\documentclass[natbib,epsfig,graphicx]{jfm}
\usepackage{amsmath}
\usepackage{bm}
\usepackage{graphicx}
\usepackage{natbib}
\usepackage{tikz}
\usetikzlibrary{shapes.geometric, arrows.meta, positioning}
\usepackage{amsmath}
\usepackage{comment}
\usepackage{subfig}
\usetikzlibrary{calc}

\usetikzlibrary{%
    arrows
}

\ifCUPmtlplainloaded \else
  \checkfont{eurm10}
  \iffontfound
    \IfFileExists{upmath.sty}
      {\typeout{^^JFound AMS Euler Roman fonts on the system,
                   using the 'upmath' package.^^J}%
       \usepackage{upmath}}
      {\typeout{^^JFound AMS Euler Roman fonts on the system, but you
                   dont seem to have the}%
       \typeout{'upmath' package installed. JFM.cls can take advantage
                 of these fonts,^^Jif you use 'upmath' package.^^J}%
      }
  \else
  \fi
\fi

\ifCUPmtlplainloaded \else
  \checkfont{msam10}
  \iffontfound
    \IfFileExists{amssymb.sty}
      {\typeout{^^JFound AMS Symbol fonts on the system, using the
                'amssymb' package.^^J}%
       \usepackage{amssymb}%
       \let\le=\leqslant  
         
      }{}
  \fi
\fi

\ifCUPmtlplainloaded \else
  \IfFileExists{amsbsy.sty}
    {\typeout{^^JFound the 'amsbsy' package on the system, using it.^^J}%
     \usepackage{amsbsy}}
    {\providecommand\boldsymbol[1]{\mbox{\boldmath $##1$}}}
\fi

\newcommand{\suchthat}{\;\ifnum\currentgrouptype=16 \middle\fi|\;}

\newsavebox{\astrutbox}
\sbox{\astrutbox}{\rule[-5pt]{0pt}{20pt}}

\title[]{Higher-order mean velocity profile in the convective atmospheric boundary layer}

\author[C.~Tong, D.~Pourabdollah, K.~Barskov and M.~Ding]%
{Chenning Tong
\thanks{Email address for correspondence: ctong@clemson.edu}
Davoud Pourabdollah, Kirill Barskov and Mengjie Ding}
\affiliation{Department of Mechanical Engineering, Clemson University, Clemson, SC 29634, USA}

\begin{document}

\maketitle

\begin{abstract}
  The higher-order mean velocity profile in the convective atmospheric boundary layer (CBL)  is derived using the method of matched asymptotic expansions.
  The universal expansion coefficients are obtained using field measurement data.
The profile  accounts for the departures from the (leading-order) log law and  local-free-convection scaling as well as the deviations from the Monin-Obukhov Similarity theory (MOST).
  Invoking MOST and the Multipoint
  Monin-Obukhov similarity theory, the perturbation equations are obtained from the Reynolds-stress, potential-temperature flux and
  potential temperature-variance budget equations and the mean momentum and  mean potential temperature equations.
  The  small parameters with the most impact in the equations are $(-z_i/L)^{-4/3}$, $(-z_i/L)^{-2/3}$ and $-h_0/L$, where $z_i$, $L$ and $h_0$ are
  the inversion height, the Obukhov length and the roughness height, respectively.
  Tong and Ding ({\it J.~Fluid Mech.} 2020) have identified the three-layer structure of the CBL with two matching (overlapping) layers, and obtained the leading-order
  expansions (the local-free-convection scaling and the log law) and the corresponding scales.
  In the present work, asymptotic matching between the outer and inner-outer layers also results in higher-order expansion terms, which account for the deviations
  from the local-free-convection scaling and the dependence on $z_i/L$.
  Asymptotic matching between the inner-outer and inner-inner layers also results in higher-order
  expansions that account for the deviations  from the log law and the dependence on $h_0/L$. The expansion coefficients are obtained
  using measurement data from the recent M$^2$HATS field campaign. Comparisons between the expansions and the measurement show  excellent agreement.
  The higher-order asymptotic expansions show that the convective logarithmic friction law derived by Tong and Ding (2020) 
  is valid to at least the second order. The predicted friction law also agrees well with measurements. The higher-order mean velocity profile can provide improved accuracy over empirical profiles.
  
\end{abstract}


\baselineskip 20pt

\newpage

\section{Introduction}

The mean velocity profile is an important property characterising the atmospheric boundary layer (ABL).  It is also important for a wide range of
practical applications such as friction drag (e.g., \citealt{KM60,TD19b}), transport of species and particulates (e.g., \citealt{TL72}), numerical weather prediction
(e.g., \citealt{Warner2010}) and wind energy (e.g., \citealt{Wolfson2012}). 
The Monin-Obukhov similarity theory (MOST,
\citealt{Obukhov46,MO54}) predicts that the non-dimensional mean shear in the surface layer depends only on $-z/L$,
where $z$ is the distance from the surface and $L=-\dfrac{u_*^3}{\kappa (g/\Theta)Q}$ is the Obukhov length. The other parameters are the friction velocity $u_*$, the von Kármán
constant $\kappa$, gravity $g$, the mean potential temperature $\Theta$, and the surface heat flux $Q$. The theory is valid asymptotically, i.e., for
$z\gg h_0$ and $z_i\gg -L$. The theory also predicts that for $h_0\ll z\ll -L$ the log law (\citealt{vK1930}) is recovered and that for $-L \ll z\ll z_i$
the local-free-convection scaling (\citealt{WCI71}) is predicted, both being the leading-order scaling.

Numerous measurements since the Kansas field campaign (e.g., \citealt{BWIB71,WCI71,KWHCICR76,Hogstrom88,Hogstrom96,Andreas2006}) have shown that the mean velocity profile
can be approximately described by  MOST.
However, the theory does not predict the dependence of  the mean profile on $z/L$, i.e., the departures from the leading-order scaling when $z\sim -L$,
which  must be determined empirically, not from the governing equations.
In addition, measurements also show deviations from the
MOST profile (i.e., the MOST-scaled profiles do not collapse) that cannot be explained by measurement uncertainties (e.g., \citealt{Hogstrom96,KB97,JSHBK01,SC12}).
The method of matched asymptotic expansions (e.g., \citealt{BO1978}), which is based on the governing equations, can be employed to systematically predict the dependence on $z/L$ and the
deviations from the MOST scaling. We provide a brief description of the method in the following.

A turbulent boundary layer can be mathematically represented as a so-called singular perturbation problem. A  shear-driven
turbulent boundary layer has two scaling layers (e.g., \citealt{Panton05}).
The boundary layer equations can be non-dimensionalised using the outer-layer parameters ($z_i$ and $u_*$) and the inner-layer parameters
($\nu$ and $u_*$ for a smooth surface) to obtain the outer and inner perturbation equations respectively, which
contain small (perturbation) parameters, such as the inverse of the friction Reynolds number.
The solution of the outer and inner equations can be written as outer and inner asymptotic expansions respectively. However, 
The outer-layer solution (the velocity defect, \citealt{vK1930}) cannot be extended to the surface while the inner-layer solution (the law of the wall, \citealt{Prandtl1925})
cannot be extended to the outer layer.
In particular, the leading-order solution for the outer equations or the inner equations (i.e., when the small parameters are set to zero) is qualitatively different from
the solution of the equations for the
entire boundary layer (i.e., when the parameters are small but non-zero), indicating that the zero-parameter point is  singular in the parameter space,
hence the term singular perturbation (\citealt{BO1978}). Asymptotic matching of the leading-order 
solutions results in the log law. Deviations from the log law in the matching layer can be accounted for using the higher-order terms in the
asymptotic expansions.

The convective boundary layer has a three layer structure (\citealt{TD19b}).
The non-dimensional boundary layer (perturbation) equations contain small parameters, such as $(-z_i/L)^{-2/3}$ and $-h_0/L$.
In the outer layer ($z \sim z_i$), which includes the mixed layer, the mean velocity profile
is in the form of the mixed-layer defect law (\citealt{TD19b}). There are two inner scaling layers, both  in the surface layer ($z\ll z_i$). In the inner-outer layer ($z\sim -L\ll z_i$),
the mean velocity profile is in the form of the surface-layer defect law (\citealt{TD19b}). This layer is the MOST layer with the leading-order non-dimensional mean shear being 
only a function of $z/L$.  In the inner-inner layer the leading-order profile is given as the law of the wall (\citealt{Prandtl1925}). 
Asymptotically matching the leading-order outer and inner-outer expansions leads to  the local-free-convection scaling, while asymptotically matching
the inner-outer and inner-inner expansions leads to the log law.
Similar to the neutral boundary layers, deviations from the leading-order scaling can be accounted for using higher-order expansions.
This method not only can identify the additional parameters that influence the mean profile, but also can quantify their influences, including the functional form. 
Therefore, higher-order asymptotic expansions provide an effective approach to systematically account for the influences of the additional parameters.
Note that $(-z_i/L)^{-2/3}$ and $-h_0/L$ are not MOST parameters, indicating that the higher-order expansions do not follow the MOST scaling, which is a leading-order prediction.
Several previous studies (e.g., \citealt{KB97,SC12})
have suggested inclusion of $z_i/L$ as an additional parameter based on dimensional analysis.
However, dimensional analysis cannot determine the relative importance of the parameters and the functional dependence of the velocity
profile on them.

In the present study, we extend the asymptotic expansion analysis of \cite{TD19b} by deriving the higher-order expansions
of the streamwise mean velocity profile in the convective boundary layer, which provide corrections
to the leading-order mean velocity profile. The analysis is based on both the MOST and the MMO scaling.
The predicted scaling of the higher-order terms are validated and the non-dimensional expansions coefficients are quantified
using field measurements from the Multipoint Monin-Obukhov Similarity Horizontal Array Turbulence Study (M$^2$HATS) field campaign (\citealt{Tong2026a}).
The higher-order expansions not only provide new insights into the structure and scaling of the CBL, but also improve the quantitative characterization of the mean velocity profile.

In Section 2, we derive the perturbation equations governing the outer, inner–outer, and inner–inner layers,
and formally express the mean velocity profile as asymptotic expansions. We then
perform higher-order asymptotic matching between the outer and inner–outer layer expansions, as well as between the inner–outer and
inner–inner layer expansions. Section 3 describes the procedures for determining the expansion coefficients using field
measurements. In Section 4, we present
the resulting expansion coefficients and validate the asymptotic expansions through comparison
with the field data. Section 5 concludes with a summary of the main findings and their implications.
Details of the procedures involved in the determination of the expansion coefficients are provided in the appendices.

\section{Perturbation equations} 

Here we obtain the perturbation equations for the three layers in the CBL. The starting point is the shear-stress budget equations, the vertical
temperature flux budget equation, the mean momentum and mean potential temperature equations, and the velocity and temperature variance budget equations (\citealt{Wyngaard2010})
\begin{equation}
  \frac{\partial{\overline{uw}}}{\partial{t}} + \overline{w^2}\frac{\partial{U}}{\partial{z}} - \frac{g}{T}\overline{u\theta} + \frac{\partial{\overline{uw^2}}}{\partial{z}} + \overline{u\frac{\partial{p}}{\partial{z}}} + \overline{w\frac{\partial{p}}{\partial{x}}} + 2(\Omega_2\overline{w^2} + \Omega_1\overline{uv} - \Omega_3\overline{vw} - \Omega_2\overline{u^2}) = 0,
  \label{eq_shear_uw}
\end{equation}
\begin{equation}
  \frac{\partial{\overline{vw}}}{\partial{t}} + \overline{w^2}\frac{\partial{V}}{\partial{z}} - \frac{g}{T}\overline{v\theta} + \frac{\partial{\overline{vw^2}}}{\partial{z}} + \overline{v\frac{\partial{p}}{\partial{z}}} + \overline{w\frac{\partial{p}}{\partial{y}}}+ 2(\Omega_1\overline{v^2} + \Omega_3\overline{uw} - \Omega_1\overline{w^2} - \Omega_2\overline{uv}) = 0,
  \label{eq_shear_vw}
\end{equation}
\begin{equation}
  \frac{\partial{\overline{w\theta}}}{\partial{t}} + \overline{w^2}\frac{\partial{\Theta}}{\partial{z}} + \frac{\partial{\overline{w^2\theta}}}{\partial{z}} + \overline{\theta\frac{\partial{p}}{\partial{z}}} + 2(\Omega_1\overline{v\theta} - \Omega_2\overline{u\theta}) + \frac{g}{T}\overline{\theta^2} = 0,
  \label{eq_flux}
\end{equation}
\begin{equation}
  \frac{\partial{\overline{vw}}}{\partial{z}} = f(U_g-U) + \frac{d\tau_{ry}}{dz},
  \label{eq_mo_vw}
\end{equation}
\begin{equation}
  \frac{\partial{\overline{uw}}}{\partial{z}} = f(V-V_g) + \frac{d\tau_{rx}}{dz},
  \label{eq_mo_uw}
\end{equation}
\begin{equation}
  \frac{\partial{\Theta}}{\partial{t}} + \frac{\partial{\overline{w\theta}}}{\partial{z}} = 0,
  \label{eq_temp}
\end{equation}
\begin{align}
\begin{split}
\frac{1}{2}\frac{\partial{\overline{w^2}}}{\partial{t}} &= -\frac{1}{2}\frac{\partial{\overline{w^3}}}{\partial{z}} + \overline{p\frac{\partial{w}}{\partial{z}}} -\overline{\frac{\partial{pw}}{\partial{z}}} + \frac{g}{T}\overline{w\theta} - {\varepsilon_3},
\label{eq_w}
\end{split}\\
\begin{split}
\frac{1}{2}\frac{\partial{\overline{u^2}}}{\partial{t}} &= -\overline{uw}\frac{\partial{U}}{\partial{z}} - \frac{1}{2}\frac{\partial{\overline{wu^2}}}{\partial{z}} + \overline{p\frac{\partial{u}}{\partial{x}}} - {\varepsilon_1},
\label{eq_u}
\end{split}\\
\begin{split}
\frac{1}{2}\frac{\partial{\overline{\theta^2}}}{\partial{t}} &= -\overline{w\theta}\frac{\partial{\Theta}}{\partial{z}} - \frac{1}{2}\frac{\partial{\overline{w\theta^2}}}{\partial{z}} - \varepsilon_\theta,
\label{eq_t}
\end{split}
\end{align}
where $\Omega_is$, $f = 2\Omega_3$, $\tau_{rx}$, $\tau_{ry}$, $\varepsilon_1$, $\varepsilon_3$, and $\varepsilon_\theta$ are the components of Earth's rotation vector,
Coriolis parameter, the shear--stress in the x- and y-directions induced by the surface roughness, dissipation rates for $\overline{u^2}/2$, $\overline{w^2}/2$, and
$\overline{\theta^2}/2$, respectively. The viscous terms except the dissipation rates in the variance budgets are neglected. The Coriolis terms in (\ref{eq_shear_uw}), (\ref{eq_shear_vw}) and (\ref{eq_flux}) are of higher order (e.g., \citealt{Wyngaard2010}), thus we neglect them in this analysis.
In the following we derive the perturbation equations in the different scaling layers.

\subsection{Outer layer}

We first derive the  the outer-layer (the mixed layer) equations. We use $U_m$, $V_g$, $z_i$, $u_*^2$, $u_*^2w_e/(fz_i)= V_gw_e$, $w_*^2$, $Q$, and $z_i/w_*$
as the outer scales for the mean velocity
components, height from the surface, the streamwise and lateral kinematic stress components, velocity variance, potential temperature flux, and time to define the
dimensionless outer variables:
\begin{align}
\begin{split}
&U(z) = U_mU_o(\frac{z}{z_i}),\ V(z) = V_gV_o(\frac{z}{z_i}),\ \Theta = \Theta_m\Theta_o(\frac{z}{z_i}),\ \overline{uw} = u_*^2\overline{uw}_o,\ \overline{vw} = \frac{u_*^2w_e}{fz_i}\overline{vw}_o,\\
&z = z_iz_o,\ \overline{u\theta} = Q\overline{u\theta}_o,\ t = \frac{z_i}{w_*}\tau,\ \overline{w^2} = w_*^2\overline{w^2_o},\ \overline{u^2} = w_*^2\overline{u_o^2},\ \overline{w\theta} = Q\overline{w\theta}_o,\ \overline{\theta^2} = (\frac{Q}{w_*})^2\overline{\theta_o^2},\\
&p = w_*^2p_o,
\end{split}\label{outer_variables}
\end{align}
where $w_*$ and $w_e$ are the mixed-layer velocity scale and the entrainment velocity (at the capping inversion),
respectively. The scales of most of the terms are obtained using MOST and the mixed-layer scaling, while the others, e.g., the velocity--pressure-gradient correlation,
are obtained using the spectral scaling
obtained using MMO (\citealt{TD19}). Substituting the outer variables given in (\ref{outer_variables}) into equations (\ref{eq_shear_uw}) to (\ref{eq_t}) we obtain
\\
\begin{equation}
\frac{\partial \overline{uw}_o}{\partial \tau}\frac{u_*^2 w_*}{z_i}
+ w_*^2 \overline{w_o^2}\frac{U_m}{z_i}\frac{\partial U_o}{\partial z_o}
- \frac{g}{T} Q \overline{u\theta}_o
+ \frac{w_* u_*^2}{z_i}
\Bigg(\frac{\partial \overline{uw^2}_o}{\partial z_o}
+ \Big(\overline{u\frac{\partial p}{\partial z}}\Big)_o
+ \Big(\overline{w\frac{\partial p}{\partial x}}\Big)_o
\Bigg)
= 0 .
\end{equation}
\begin{equation}
\frac{\partial{\overline{vw}_o}}{\partial{\tau}}\frac{u_*^2w_e}{fz_i}\frac{w_*}{z_i} + w_*^2\overline{w^2_o}\frac{V_g}{z_i}\frac{\partial{V_o}}{\partial{z_o}} - \frac{g}{T}Q\overline{v\theta}_o + \frac{w_*}{z_i}\frac{u_*^2w_e}{fz_i}\Big(\frac{\partial{\overline{vw^2}_o}}{\partial{z_o}} + \Big(\overline{v\frac{\partial{p}}{\partial{z}}}\Big)_o + \Big(\overline{w\frac{\partial{p}}{\partial{y}}}\Big)_o \Big) = 0,
\end{equation}
\begin{equation}
\frac{\partial{\overline{w\theta}_o}}{\partial{\tau}}\frac{Qw_*}{z_i} + w_*^2\overline{w_o^2}\frac{\partial{\Theta_o}}{\partial{z_o}}\frac{\Theta_m}{z_i} + \frac{Qw_*}{z_i}\Big(\frac{\partial{\overline{w^2\theta_o}}}{\partial{z_o}} + \Big(\overline{\theta\frac{\partial{p}}{\partial{z}}}\Big)_o\Big) + \frac{g}{T}\Big(\frac{Q}{w_*}\Big)^2\overline{\theta_o^2} = 0,
\end{equation}
\begin{equation}
\frac{u_*^2}{z_i} \frac{\partial{\overline{uw}_o}}{\partial{z_o}} = fV_g ( V_o - 1),
\end{equation}
\begin{equation}
\frac{u_*^2w_e}{fz_i} \frac{1}{z_i} \frac{\partial{\overline{vw}_o}}{\partial{z_o}} = fU_m \frac{U_g - U}{U_m},\ \text{with}\  U_m \equiv U_g - \frac{u_*^2w_e}{f^2 z_i^2}.
\end{equation}
\begin{equation}
\frac{\Theta_m}{z_i/w_*} \frac{\partial{\Theta_o}}{\partial{\tau}} + \frac{Q}{z_i} \frac{\partial{\overline{w\theta_o}}}{\partial{z_o}} = 0,
\end{equation}
\begin{equation}
\frac{1}{2} \frac{w_*^2}{z_i/w_*} \frac{\partial{\overline{w_o^2}}}{\partial{\tau}} = -\frac{1}{2} \frac{w_*^3}{z_i} \frac{\partial{\overline{w_o^3}}}{\partial{z_o}} + \frac{w_*^3}{z_i} \Big(\overline{p\frac{\partial{w}}{\partial{z}}}\Big)_o - \frac{w_*^3}{z_i} \Big(\overline{\frac{\partial{pw}}{\partial{z}}}\Big)_o + \frac{g}{T}Q\overline{w\theta}_o - \varepsilon_{3},
\end{equation}
\begin{equation}
\frac{1}{2} \frac{w_*^2}{z_i/w_*} \frac{\partial{\overline{u_o^2}}}{\partial{\tau}} = \frac{u_*^3}{z_i}\Big(-\frac{z_i}{L}\Big)^{-1/3} \overline{uw}_o\Big(\frac{\partial{U}}{\partial{z}}\Big)_o - \frac{1}{2} \frac{w_*^3}{z_i} \frac{\partial{\overline{wu_o^2}}}{\partial{z_o}} + \frac{w_*^3}{z_i} \Big(\overline{p\frac{\partial{u}}{\partial{x}}}\Big)_o - \varepsilon_{1},
\end{equation}
\begin{equation}
\frac{1}{2} \Big(\frac{Q}{w_*}\Big)^2 \frac{w_*}{z_i} \frac{\partial{\overline{\theta_o^2}}}{\partial{\tau}} = -Q \frac{Q}{w_*z_i} \overline{w\theta}_o\Big(\frac{\partial{\Theta}}{\partial{z}}\Big)_o - \frac{1}{2} \Big(\frac{Q}{w_*}\Big)^2 \frac{w_*}{z_i} \frac{\partial{\overline{w\theta^2}_o}}{\partial{z_o}} - \varepsilon_{{\theta}}.
\end{equation}
\\
Using $w_*^2U_m/z_i$,  $Qw_*/z_i$, $u_*^2/z_i$, $\Theta_m w_*^2/z_i$,  $w_*^3/z_i$ and $Q^2/w_*z_i$ to non-dimensionalise the shear-stress and vertical flux budgets, the mean momentum and
mean potential temperature equations, and the variance budgets respectively, we obtain the outer equations
\begin{equation}
  \epsilon_1 \frac{\partial{\overline{uw}_o}}{\partial{\tau}} + \overline{w_o^2}\frac{\partial{U_o}}{\partial{z_o}} - \frac{w_*}{U_m}\overline{u\theta_o} + \epsilon_1 \Big( \frac{\partial{\overline{uw_o^2}}}{\partial{z_o}} + \Big(\overline{u\frac{\partial{p}}{\partial{z}}}\Big)_o + \Big(\overline{w\frac{\partial{p}}{\partial{x}}}\Big)_o \Big) = 0,
  \label{eq:o_uw}
\end{equation}
\begin{equation}
\epsilon_2 \frac{V_g}{U_m} \frac{\partial{\overline{vw}_o}}{\partial{\tau}} + \overline{w_o^2} \frac{V_g}{U_m} \frac{\partial{V_o}}{\partial{z_o}} - \frac{w_*}{U_m}\overline{v\theta_o} + \epsilon_2 \frac{V_g}{U_m} \Big( \frac{\partial{\overline{vw_o^2}}}{\partial{z_o}} + \Big(\overline{v\frac{\partial{p}}{\partial{z}}}\Big)_o + \Big(\overline{w\frac{\partial{p}}{\partial{y}}}\Big)_o\Big) = 0,
\end{equation}
\begin{equation}
\epsilon_{\theta} \frac{\partial{\overline{w\theta}_o}}{\partial{\tau}} + \overline{w_o^2}\frac{\partial{\Theta_o}}{\partial{z_o}} + \epsilon_{\theta} \Big(\frac{\partial{\overline{w^2\theta_o}}}{\partial{z_o}} + \Big(\overline{\theta\frac{\partial{p}}{\partial{z}}}\Big)_o\Big) + \epsilon_{\theta} \overline{\theta_o^2} = 0,
\end{equation}
\begin{equation}
\frac{\partial{\overline{uw}_o}}{\partial{z_o}} =-V_o + 1 = 1 - \epsilon_2 V_{o,2},
\end{equation}
\begin{equation}
\frac{\partial{\overline{vw_o}}}{\partial{z_o}} = 1 - \epsilon_4 U_{o,2}.
\end{equation}
\begin{equation}
\frac{\partial{\Theta_o}}{\partial{\tau}} + \epsilon_{\theta} \frac{\partial{\overline{w\theta_o}}}{\partial{z_o}} = 0,
\end{equation}
\begin{equation}
  \frac{1}{2}\frac{\partial{\overline{w_o^2}}}{\partial{\tau}} = -\frac{1}{2}\frac{\partial{\overline{w_o^3}}}{\partial{z_o}} + \Big(\overline{p\frac{\partial{w}}{\partial{z}}}\Big)_o - \Big(\overline{\frac{\partial{pw}}{\partial{z}}}\Big)_o + \overline{w\theta}_o - \varepsilon_{3o},
  \label{eq:o_w2}
\end{equation}
\begin{equation}
  \frac{1}{2}\frac{\partial{\overline{u_o^2}}}{\partial{\tau}} = -\epsilon_3\overline{uw}_o\Big(\frac{\partial{U}}{\partial{z}}\Big)_o - \frac{1}{2}\frac{\partial{\overline{wu_o^2}}}{\partial{z_o}} + \Big(\overline{p\frac{\partial{u}}{\partial{x}}}\Big)_o - \varepsilon_{1o},
  \label{eq:o_u2}
\end{equation}
\begin{equation}
  \frac{1}{2}\frac{\partial{\overline{\theta_o^2}}}{\partial{\tau}} = -\overline{w\theta}_o\Big(\frac{\partial{\Theta}}{\partial{z}}\Big)_o - \frac{1}{2}\frac{\partial{\overline{w\theta^2}_o}}{\partial{z_o}} - \varepsilon_{{\theta}o}
\end{equation}
The small parameters in the outer equations are
\begin{equation}
\epsilon_1 = \frac{u_*}{w_*}\frac{u_*}{U_m},\  \epsilon_2 = \frac{w_e}{w_*},\  \epsilon_3 = (-\frac{z_i}{L})^{-4/3},\  \epsilon_4 = \epsilon_1 U_m \Big/ \frac{V_gw_e}{fz_i},\ \epsilon_{\theta} = \frac{Q}{w_*\Theta_m}.
\end{equation}
Here $\epsilon_1$ is the relative scale of the velocity defect to the mean velocity in the outer layer and 
$\epsilon_3$ quantifies the influence of the shear production of $\overline {u^2}$. In the outer layer, $z>-L$, therefore the shear production is a small term.
$\epsilon_2$ and $\epsilon_4$ quantify the influence of entrainment and the Coriolis force on
the $U-$ and $V$-components of the mean velocity. $\epsilon_\theta$ is the relative scale of the potential temperature defect relative to
the difference of the mixed-layer mean temperature and surface mean temperature.
With these small parameters, the outer solution can be written as the following outer asymptotic expansions 
\begin{equation}
  U_o = U_{o,1} + \epsilon_1 U_{o,2} + \epsilon_1\epsilon_3 U_{o,31} + \epsilon_1\epsilon_2 U_{o,32} + \cdot\cdot\cdot,
  \label{eq:Uo}
\end{equation}
\begin{equation}
\overline{w_o^2}(z_o) = \overline{w_o^2}_{,1}(z_o) + \epsilon_3 \overline{w^2_{o,2}}(z_o) + O(\epsilon_3^2),
\end{equation}
\begin{equation}
\frac{\partial{\overline{uw^2}_{o}}}{\partial{z_o}} = \frac{\partial{\overline{uw^2}_{o,1}}}{\partial{z_o}} + \epsilon_3\frac{\partial{\overline{uw^2}_{o,21}}}{\partial{z_o}} + \epsilon_2\frac{\partial{\overline{uw^2}_{o,22}}}{\partial{z_o}} + \cdot\cdot\cdot,
\end{equation}
\begin{equation}
\frac{\partial{\overline{uw}_{o}}}{\partial{z_o}} = \frac{\partial{\overline{uw}_{o,1}}}{\partial{z_o}} + \epsilon_3\frac{\partial{\overline{uw}_{o,21}}}{\partial{z_o}} + \epsilon_2\frac{\partial{\overline{uw}_{o,22}}}{\partial{z_o}} + \cdot\cdot\cdot,
\end{equation}
\begin{equation}
V_o = V_{o,1} + \epsilon_2 V_{o,2} + \epsilon_2\epsilon_3 V_{o,31} + \epsilon_2\epsilon_4 V_{o,32} + \cdot\cdot\cdot,
\end{equation}
\begin{equation}
\frac{\partial{\overline{vw^2}_{o}}}{\partial{z_o}} = \frac{\partial{\overline{vw^2}_{o,1}}}{\partial{z_o}} + \epsilon_3\frac{\partial{\overline{vw^2}_{o,21}}}{\partial{z_o}} + \epsilon_4\frac{\partial{\overline{vw^2}_{o,22}}}{\partial{z_o}} + \cdot\cdot\cdot,
\end{equation}
\begin{equation}
\frac{\partial{\overline{vw}_{o}}}{\partial{z_o}} = \frac{\partial{\overline{vw}_{o,1}}}{\partial{z_o}} + \epsilon_3\frac{\partial{\overline{vw}_{o,21}}}{\partial{z_o}} + \epsilon_4\frac{\partial{\overline{vw}_{o,22}}}{\partial{z_o}} + \cdot\cdot\cdot.
\end{equation}
\begin{equation}
\Theta_o = \Theta_{o,1} + \epsilon_{\theta} \Theta_{o,2} + \epsilon_{\theta}\epsilon_3 \Theta_{o,3} + \cdot\cdot\cdot,
\end{equation}
Substituting the outer expansions back to equations (\ref{eq:o_uw}), (\ref{eq:o_w2}) and (\ref{eq:o_u2}) we obtain
\begin{align}
\begin{split}
  &\epsilon_1 \frac{\partial{\overline{uw}_o}}{\partial{\tau}} + \Big(\overline{w_{o,1}^2} + \epsilon_3\overline{w_{o,2}^2} + \cdot\cdot\cdot\Big) \Big(\frac{\partial{U_{o,1}}}{\partial{z_o}} + \epsilon_1 \frac{\partial{U_{o,2}}}{\partial{z_o}} + \epsilon_1\epsilon_3 \frac{\partial{U_{o,31}}}{\partial{z_o}} + \epsilon_1\epsilon_2 \frac{\partial{U_{o,32}}}{\partial{z_o}} + \cdot\cdot\cdot\Big)\\
  & + \cdot\cdot\cdot  + \epsilon_1\Big(\Big(\overline{w\frac{\partial{{p}}}{\partial{x}}}\Big)_{o,1} + \epsilon_3\Big(\overline{w\frac{\partial{{p}}}{\partial{x}}}\Big)_{o,2} + \epsilon_2\Big(\overline{w\frac{\partial{{p}}}{\partial{x}}}\Big)_{o,3} + \cdot\cdot\cdot\Big) = 0.
\end{split}
\end{align}

\begin{align}
\begin{split}
\frac{1}{2}\Bigg(
\frac{\partial \overline{w^2_{o,1}}}{\partial \tau}
+ \epsilon_3 \frac{\partial \overline{w^2_{o,2}}}{\partial \tau} + \cdot\cdot\cdot
\Bigg)
&= -\frac{1}{2}\Bigg(
\frac{\partial \overline{w^3_{o,1}}}{\partial z_o}
+ \epsilon_3 \frac{\partial \overline{w^3_{o,2}}}{\partial z_o} + \cdot\cdot\cdot
\Bigg) + \cdot\cdot\cdot\\
& + \Big(\overline{w\theta}_{o,1} + \epsilon_\theta \overline{w\theta}_{o,2} + \cdot\cdot\cdot\Big) - \varepsilon_{3o}
\end{split}
\end{align}

\begin{align}
\begin{split}
\frac{1}{2}\frac{\partial \overline{u^2_o}}{\partial \tau}
&= - \epsilon_3 \Big(\overline{uw}_{o,1} + \epsilon_3\overline{uw}_{o,2} + \epsilon_2\overline{uw}_{o,3} + \cdot\cdot\cdot\Big)\Big(\frac{\partial{U_{o,1}}}{\partial{z_o}} + \epsilon_1 \frac{\partial{U_{o,2}}}{\partial{z_o}} + \epsilon_1\epsilon_3 \frac{\partial{U_{o,31}}}{\partial{z_o}}\\
& + \epsilon_1\epsilon_2 \frac{\partial{U_{o,32}}}{\partial{z_o}} + \cdot\cdot\cdot\Big) - \frac{1}{2}\Big(\frac{\partial \overline{uw^2}_{o,1}}{\partial z_o}
+ \epsilon_3 \frac{\partial \overline{uw^2}_{o,21}}{\partial z_o} + \epsilon_2 \frac{\partial \overline{uw^2}_{o,22}}{\partial z_o} + \cdot\cdot\cdot\Big) + \cdot\cdot\cdot - \varepsilon_{1o}
\end{split}
\end{align}
It can be seen that the second-order mean velocity gradient is induced by the second-order pressure--velocity-gradient interaction term, which is in turn induced by the
mean shear production of $\overline{u_{o}^2}$. The last term is of second order in the outer layer, but becomes leading order for $-z/L\sim 1$, and therefore is
the source of the singularity for the outer equations.

\begin{table} 
\begin{center} 
\begin{tabular}{c  c c c c c c} 
\hline 
        & $\epsilon_1$   & $\epsilon_2$ & $\epsilon_3$ & $\epsilon_4$ & \bf{--} & $\epsilon_{\theta}$ \\ 
 Outer  &  & & & &  &  \\
 layer  & $\displaystyle \frac{u_*}{w_*}\frac{u_*}{U_m}$  & $\displaystyle \frac{w_e}{w_*}$ & $\displaystyle (-\frac{z_i}{L})^{-\frac{4}{3}}$ & $\displaystyle \frac{\epsilon_1 U_m}{{V_gw_e}/{fz_i}}$ & \bf{--}  & $\displaystyle \frac{Q}{w_* \Theta_m}$ \\ 
\hline
  & $\epsilon_1'$   & $\epsilon_2'$ & $\epsilon_3'$ & $\epsilon_4'$ & $\epsilon_5'$ & $\epsilon_{\theta}'$ \\ 
 Inner-outer  &  & & & &  &  \\
 layer  & $\displaystyle \frac{u_*}{U_m}$   & $\displaystyle \frac{u_*^2}{w_*^2}\frac{w_e}{w_*}$ & $\displaystyle (-\frac{z_i}{L})^{-\frac{2}{3}}$ & $\displaystyle \frac{\epsilon_1'U_m}{{V_gw_e}/{fz_i}}$ & $\displaystyle \frac{L}{z_i}$ & $\displaystyle \frac{Q}{u_*\Theta_m}$ \\ 
\hline 
  & $\epsilon_1'$   & $\epsilon_2''$ & $\epsilon_3''$ & $\epsilon_4''$ & $\epsilon_5''$ & $\epsilon_{\theta}'$ \\ 
 Inner-inner  &  & & & &  &  \\
 layer  & $\displaystyle \frac{u_*}{U_m}$   & $\displaystyle \frac{fh_0}{u_*}\frac{U_g}{V_g}$ & $\displaystyle \frac{h_0}{L}$ & $\displaystyle \frac{u_*}{U_g}$ & $\displaystyle \frac{h_0}{z_i}$ & $\displaystyle \frac{Q}{u_*\Theta_m}$ \\ 
\hline 
  
\end{tabular} 
\caption{Small parameters in the perturbation equations. } 
\label{tab:small} 
\end{center} 
\end{table}

\subsection{Inner-outer layer}

Due to the  mean shear production term being of leading order for $z \sim -L$, the leading-order outer solution is no longer valid, indicating that there exists
an inner scaling layer
there (\citealt{TD19b}). In fact, 
there are two inner layers, the inner-outer layer with a thickness of $-L$ and the inner-inner layer with a thickness of $h_0$.
We define the non-dimensional inner-outer variables as
\begin{align}
\begin{split}
  &U(z) = U_mU_{io}(-\frac{z}{L}),\ V(z) = V_gV_{io}(-\frac{z}{L}),\ \Theta = \Theta_m\Theta_{io}(-\frac{z}{L}),\ \overline{uw} = u_*^2\overline{uw}_{io},\\
  &\overline{vw} = \frac{u_*^2w_e}{fz_i}(-\frac{L}{z_i})\overline{vw}_{io},\ \
  z = -Lz_{io},\ \overline{u\theta} = Q\overline{u\theta}_{io},\ t = \frac{z_i}{w_*}\tau,\ \overline{w^2} = u_*^2\overline{w^2_{io}},\ \overline{u^2} = u_*^2\overline{u_{io}^2},\\
  &\overline{w\theta} = Q\overline{w\theta}_{io},\ \overline{\theta^2} = (\frac{Q}{u_*})^2\overline{\theta_{io}^2},\ \ p = u_*^2p_{io},
\end{split}\label{inner_outer_variables}
\end{align}
Substituting the variables given in (\ref{inner_outer_variables}) into equations (\ref{eq_shear_uw}) to (\ref{eq_t}) we obtain  
\begin{equation}
\frac{u_*^2w_*}{z_i} \frac{\partial{\overline{uw}_{io}}}{\partial{\tau}} + u_*^2\overline{w^2_{io}}\frac{U_m}{-L}\frac{\partial{U_{io}}}{\partial{z_{io}}} - \frac{g}{T}Q\overline{u\theta}_{io} + \frac{u_*^3}{-L}\Big(\frac{\partial{\overline{uw^2}_{io}}}{\partial{z_{io}}} + \Big(\overline{w\frac{\partial{p}}{\partial{x}}}\Big)_{io} + \Big(\overline{u\frac{\partial{p}}{\partial{z}}}\Big)_{io}\Big) = 0,
\end{equation}
\begin{equation}
\frac{u_*^2w_e}{fz_i}(-\frac{L}{z_i})\frac{w_*}{z_i} \frac{\partial{\overline{vw}_{io}}}{\partial{\tau}} + u_*^2\overline{w^2_{io}}\frac{V_g}{-L}\frac{\partial{V_{io}}}{\partial{z_{io}}} - \frac{g}{T}Q\overline{v\theta}_{io} + \frac{u_*}{z_i}\frac{u_*^2w_e}{fz_i}\Big(\frac{\partial{\overline{vw^2}_{io}}}{\partial{z_{io}}} + \Big(\overline{w\frac{\partial{p}}{\partial{y}}}\Big)_{io} + \Big(\overline{v\frac{\partial{p}}{\partial{z}}}\Big)_{io}\Big) = 0,
\end{equation}
\begin{equation}
\frac{Qw_*}{z_i} \frac{\partial{\overline{w\theta}_{io}}}{\partial{\tau}} + u_*^2\overline{w^2_{io}}\frac{\Theta_m}{-L}\frac{\partial{\Theta_{io}}}{\partial{z_{io}}} + \frac{g}{T}\Big(\frac{Q}{u_*}\Big)^2\overline{\theta_{io}^2} + \frac{Qu_*}{-L}\Big(\frac{\partial{\overline{w^2\theta}_{io}}}{\partial{z_{io}}} + \Big(\overline{\theta\frac{\partial{p}}{\partial{z}}}\Big)_{io}\Big) = 0,
\end{equation}
\begin{equation}
\frac{u_*^2}{L} \frac{\partial{\overline{uw_{io}}}}{\partial{z_{io}}} = f(V-V_g),
\end{equation}
\begin{equation}
\frac{u_*^2w_e}{fz_i} \Big(-\frac{L}{z_i}\Big) \frac{1}{L} \frac{\partial{\overline{vw_{io}}}}{\partial{z_{io}}} = f(U_g - U). 
\end{equation}
\begin{equation}
\frac{\Theta_m}{z_i/w_*} \frac{\partial{\Theta_{io}}}{\partial{\tau}} + \frac{Q}{L} \frac{\partial{\overline{w\theta_{io}}}}{\partial{z_{io}}} = 0,
\end{equation}
\begin{align}
\begin{split}
\frac{1}{2}\frac{u_*^2w_*}{z_i}\frac{\partial{\overline{w_{io}^2}}}{\partial{\tau}} &= -\frac{1}{2}\frac{\partial{\overline{w_{io}^3}}}{\partial{z_{io}}}\frac{u_*^3}{L} + \Bigg(\overline{p\frac{\partial{w}}{\partial{z}}}\Bigg)_{io}\frac{u_*^3}{L} - \Bigg(\overline{\frac{\partial{pw}}{\partial{z}}}\Bigg)_{io}\frac{u_*^3}{L} + \frac{g}{T}Q\overline{w\theta}_{io} - {\varepsilon_{3}},
\end{split}\\
\begin{split}
\frac{1}{2}\frac{u_*^2w_*}{z_i}\frac{\partial{\overline{u_{io}^2}}}{\partial{\tau}} &= -\overline{uw}_{io}\bigg(\frac{\partial{U}}{\partial{z}}\bigg)_{io}\frac{u_*^3}{L} - \frac{1}{2}\frac{\partial{\overline{wu_{io}^2}}}{\partial{z_{io}}}\frac{u_*^3}{L} + \Bigg(\overline{p\frac{\partial{u}}{\partial{x}}}\Bigg)_{io}\frac{u_*^3}{L} - {\varepsilon_{1}},
\end{split}
\\
\begin{split}
\frac{1}{2}\bigg(\frac{Q}{u_*}\bigg)^2\frac{w_*}{z_i}\frac{\partial{\overline{\theta_{io}^2}}}{\partial{\tau}} &= -Q\frac{Q}{u_*L}\overline{w\theta}_{io}\Big(\frac{\partial{\Theta}}{\partial{z}}\Big)_{io} -\frac{1}{2}\bigg(\frac{Q}{u_*}\bigg)^2\frac{u_*}{L}\frac{\partial{\overline{w\theta_{io}^2}}}{\partial{z_{io}}} - \varepsilon_{{\theta}}.
\end{split}
\end{align}
\\
Using $u_*^2U_m/(-L)$,  $Qu_*/(-L)$, $u_*^2/(-L)$, $\Theta_m u_*^2/(-L)$,  $u_*^3/(-L)$ and $Q^2/u_*(-L)$ to non-dimensionalise the shear-stress and vertical flux budgets, the mean momentum and mean potential temperature equations, as well as the associated variance budgets, we obtain the corresponding inner–outer equations
\begin{equation}
  \epsilon_1' \epsilon_3' \frac{\partial{\overline{uw}_{io}}}{\partial{\tau}} + \overline{w^2_{io}}\frac{\partial{U_{io}}}{\partial{z_{io}}} - \epsilon_1' \overline{u\theta}_{io} + \epsilon_1' \Big(\frac{\partial{\overline{uw^2}_{io}}}{\partial{z_{io}}} + \Big(\overline{w\frac{\partial{p}}{\partial{x}}}\Big)_{io} + \Big(\overline{u\frac{\partial{p}}{\partial{z}}}\Big)_{io}\Big) = 0,
  \label{eq:io_uw}
\end{equation}
\begin{equation}
\frac{u_*^4w_e}{w_*^5}\frac{V_g}{U_m}\frac{\partial{\overline{vw}_{io}}}{\partial{\tau}} + \overline{w^2_{io}}\frac{V_g}{U_m}\frac{\partial{V_{io}}}{\partial{z_{io}}} - \epsilon_1' \overline{v\theta}_{io} + \epsilon_2' \frac{V_g}{U_m}\Big(\frac{\partial{\overline{vw^2}_{io}}}{\partial{z_{io}}} + \Big(\overline{w\frac{\partial{p}}{\partial{y}}}\Big)_{io} + \Big(\overline{v\frac{\partial{p}}{\partial{z}}}\Big)_{io}\Big) = 0.
\end{equation}
\begin{equation}
\epsilon_3' \epsilon_{\theta}' \frac{\partial{\overline{w\theta}_{io}}}{\partial{\tau}} + \overline{w^2_{io}}\frac{\partial{\Theta_{io}}}{\partial{z_{io}}} + \epsilon_{\theta}' \overline{\theta_{io}^2} + \epsilon_{\theta}' \Big(\frac{\partial{\overline{w^2\theta_{io}}}}{\partial{z_{io}}} + \Big(\overline{\theta\frac{\partial{p}}{\partial{z}}}\Big)_{io}\Big) = 0,
\end{equation}
\begin{equation}
\frac{\partial{\overline{uw_{io}}}}{\partial{z_{io}}} = -\epsilon_5' - \epsilon_5' \epsilon_2' V_{io,2}
\end{equation}
\begin{equation}
\frac{\partial{\overline{vw_{io}}}}{\partial{z_{io}}} = 1 - \epsilon_4' U_{io,2}
\end{equation}
\begin{equation}
\epsilon_3' \frac{\partial{\Theta_{io}}}{\partial{\tau}} + \epsilon_{\theta}' \frac{\partial{\overline{w\theta_{io}}}}{\partial{z_{io}}} = 0,
\end{equation}
\begin{align}
\begin{split}
\frac{1}{2}\epsilon_3'\frac{\partial{\overline{w_{io}^2}}}{\partial{\tau}} &= -\frac{1}{2}\frac{\partial{\overline{w_{io}^3}}}{\partial{z_{io}}} + \Big(\overline{p\frac{\partial{w}}{\partial{z}}}\Big)_{io} - \Big(\overline{\frac{\partial{pw}}{\partial{z}}}\Big)_{io} + \overline{w\theta}_{io} - {\varepsilon_{3io}},
\label{eq_io_w2}
\end{split}\\
\begin{split}
\frac{1}{2}\epsilon_3'\frac{\partial{\overline{u_{io}^2}}}{\partial{\tau}} &= -\overline{uw}_{io}\Big(\frac{\partial{U}}{\partial{z}}\Big)_{io} - \frac{1}{2}\frac{\partial{\overline{wu^2_{io}}}}{\partial{z_{io}}} + \Big(\overline{p\frac{\partial{u}}{\partial{x}}}\Big)_{io} - {\varepsilon_{1io}},
\label{eq_io_u2}
\end{split}\\
\begin{split}
\frac{1}{2}\epsilon_3'\frac{\partial{\overline{\theta_{io}^2}}}{\partial{\tau}} &= -\overline{w\theta}_{io}\frac{\partial{\Theta_{io}}}{\partial{z_{io}}} -\frac{1}{2}\frac{\partial{\overline{w\theta^2_{io}}}}{\partial{z_{io}}} - \varepsilon_{{\theta}io},
\label{eq_io_theta2}
\end{split}
\end{align}
where
\begin{align}
\epsilon_1' = \frac{u_*}{U_m},\ 
\epsilon_2' = \frac{u_*^2}{w_*^2}\frac{w_e}{w_*},\ 
\epsilon_3' = -\frac{L}{z_i}\frac{w_*}{u_*} = {\kappa}^{-\frac{1}{3}}(-\frac{z_i}{L})^{-\frac{2}{3}},\
\epsilon_4' = \frac{U_m}{V_gw_e/fz_i} \epsilon_1',\ 
\epsilon_5'= \frac{L}{z_i},\ 
\epsilon_{\theta}' = \frac{Q}{u_*\Theta_m}.
\end{align}
Here $\epsilon'_1$ is the scale of the velocity defect in the inner-outer layer relative to the mixed-layer mean velocity scale.
$\epsilon'_2$ and $\epsilon'_4$ quantify the influence of the entrainment. $\epsilon'_3$ quantifies the influence of the
non-stationarity. $\epsilon'_5$ quantifies the effect of the variations of the shear stress
with height (the shear stress is treated as a constant only at the leading order). $\epsilon'_\theta$ represents the relative scale of the temperature defect.

The inner-outer expansions are
\begin{equation}
\overline{w_{io}^2}(z_{io}) = \overline{w_{io}^2}_{,1}(z_{io}) + \epsilon_3' \overline{w^2_{io,2}(z_{io})} + O(\epsilon_3''^2),
\end{equation}
\begin{equation}
  U_{io} = U_{io,1} + \epsilon_1' U_{io,2} + \epsilon_1' \epsilon_3' U_{io,31} + \epsilon_1' \epsilon_5' U_{io,32} + \cdot\cdot\cdot
  \label{eq:Uio}
\end{equation}
\begin{equation}
\frac{\partial{\overline{uw_{io}^2}}}{\partial{z_{io}}} = \frac{\partial{\overline{uw_{io,1}^2}}}{\partial{z_{io}}} + \epsilon_3' \frac{\partial{\overline{uw_{io,2}^2}}}{\partial{z_{io}}} + \epsilon_5' \epsilon_2' \frac{\partial{\overline{uw_{io,3}^2}}}{\partial{z_{io}}} + \cdot\cdot\cdot
\end{equation}
\begin{equation}
\frac{\partial{\overline{uw_{io}}}}{\partial{z_{io}}} = \frac{\partial{\overline{uw_{io,1}}}}{\partial{z_{io}}} + \epsilon_3' \frac{\partial{\overline{uw_{io,2}}}}{\partial{z_{io}}} + \epsilon_5' \epsilon_2' \frac{\partial{\overline{uw_{io,3}}}}{\partial{z_{io}}} + \cdot\cdot\cdot
\end{equation}
\begin{equation}
V_{io} = V_{io,1} + \epsilon_2' V_{io,2} + \epsilon_2' \epsilon_3' V_{io,31} + \epsilon_2' \epsilon_4' V_{io,32} + \cdot\cdot\cdot
\end{equation}
\begin{equation}
\frac{\partial{\overline{vw_{io}^2}}}{\partial{z_{io}}} = \frac{\partial{\overline{vw_{io,1}^2}}}{\partial{z_{io}}} + \epsilon_3' \frac{\partial{\overline{vw_{io,21}^2}}}{\partial{z_{io}}} + \epsilon_4' \frac{\partial{\overline{vw_{io,22}^2}}}{\partial{z_{io}}} + \cdot\cdot\cdot
\end{equation}
\begin{equation}
\frac{\partial{\overline{vw_{io}}}}{\partial{z_{io}}} = \frac{\partial{\overline{vw_{io,1}}}}{\partial{z_{io}}} + \epsilon_3' \frac{\partial{\overline{vw_{io,21}}}}{\partial{z_{io}}} + \epsilon_4' \frac{\partial{\overline{vw_{io,22}}}}{\partial{z_{io}}} + \cdot\cdot\cdot
\end{equation}
\begin{equation}
\Theta_{io} = \Theta_{io,1} + \epsilon_{\theta}' \Theta_{io,2} + \epsilon_{\theta}'\epsilon_3' \Theta_{io,3} + \cdot\cdot\cdot,
\end{equation}
Substitute the inner-outer expansions back to equations (\ref{eq:io_uw}), (\ref{eq_io_w2}) and (\ref{eq_io_u2}) we obtain 
\begin{align}
\begin{split}
&\epsilon_1' \epsilon_3'\frac{\partial{\overline{uw}_{io}}}{\partial{\tau}} +\Big(\overline{w_{io,1}^2} + \epsilon_3' \overline{w^2_{io,2}} + \cdot\cdot\cdot\Big) \Big(\frac{\partial{U_{io,1}}}{\partial{z_{io}}} + \epsilon_1' \frac{\partial{U_{io,2}}}{\partial{z_{io}}} + \epsilon_1' \epsilon_3' \frac{\partial{U_{io,31}}}{\partial{z_{io}}} + \epsilon_1' \epsilon_5' \frac{\partial{U_{io,32}}}{\partial{z_{io}}} + \cdot\cdot\cdot\Big) \\
  & + \epsilon_1'\Big(\Big(\overline{w\frac{\partial{{p}}}{\partial{x}}}\Big)_{io,1} + \epsilon_3'\Big(\overline{w\frac{\partial{{p}}}{\partial{x}}}\Big)_{io,2} + \epsilon_5' \epsilon_2'\Big(\overline{w\frac{\partial{{p}}}{\partial{x}}}\Big)_{io,3}\Big) + \cdot\cdot\cdot = 0.
\end{split}
\end{align}

\begin{align}
\begin{split}
\frac{1}{2}\epsilon_3'\Bigg(
\frac{\partial \overline{w^2_{io,1}}}{\partial \tau}
+ \epsilon_3' \frac{\partial \overline{w^2_{io,2}}}{\partial \tau} + \cdot\cdot\cdot
\Bigg)
&= -\frac{1}{2}\Bigg(
\frac{\partial \overline{w^3_{io,1}}}{\partial z_{io}} + \epsilon_3' \frac{\partial \overline{w^3_{io,2}}}{\partial z_{io}} + \cdot\cdot\cdot\Bigg) + \cdot\cdot\cdot\\
& + \Big(\overline{w\theta}_{io,1} + \epsilon_\theta \overline{w\theta}_{io,2} + \cdot\cdot\cdot\Big) - \varepsilon_{3io}
\end{split}
\end{align}

\begin{align}
\begin{split}
\frac{1}{2}\epsilon_3'\frac{\partial \overline{u^2_{io}}}{\partial \tau}
&= - \Big(\overline{uw}_{io,1} + \epsilon_3'\overline{uw}_{io,2} + \epsilon_5' \epsilon_2'\overline{uw}_{io,3} + \cdot\cdot\cdot\Big)\Big(\frac{\partial{U_{io,1}}}{\partial{z_{io}}} + \epsilon_1' \frac{\partial{U_{io,2}}}{\partial{z_{io}}}\\
& + \epsilon_1'\epsilon_3' \frac{\partial{U_{io,31}}}{\partial{z_{io}}}
 + \epsilon_1'\epsilon_5' \frac{\partial{U_{io,32}}}{\partial{z_{io}}} + \cdot\cdot\cdot\Big)
- \frac{1}{2}\Big(
\frac{\partial \overline{uw^2}_{io,1}}{\partial z_{io}}
+ \epsilon_3' \frac{\partial \overline{uw^2}_{io,21}}{\partial z_{io}}\\
& + \epsilon_5' \epsilon_2' \frac{\partial \overline{uw^2}_{io,22}}{\partial z_{io}} + \cdot\cdot\cdot\Big) + \cdot\cdot\cdot - \varepsilon_{1io}
\end{split}
\end{align}
The time derivative terms are of second order in the inner-outer layer and become of leading order in the outer layer.
Therefore, the leading-order inner-outer solution is not valid for $z\sim z_i$. These terms are a source of the second-order
velocity gradient.

\subsection{Inner-inner layer}

The leading-order inner-outer solution is not valid for $z \sim h_0$, indicating that there is
a second inner scaling layer, the roughness layer, which is nested inside the inner-outer layer, which we call the inner-inner layer (\citealt{TD19b}). Therefore, we also need to
obtain the inner-inner expansions.
We define the non-dimensional inner-inner variables as
\begin{align}
\begin{split}
&U(z) = U_mU_{ii}(\frac{z}{h_0}),\ V(z) = V_gV_{ii}(\frac{z}{h_0}),\ \Theta = \Theta_m\Theta_{ii}(\frac{z}{h_0}),\ \overline{uw} = u_*^2\overline{uw}_{ii},\ \overline{vw} = fU_gh_0\overline{vw}_{ii},\\
&z = h_0z_{ii},\ \overline{u\theta} = Q\overline{u\theta}_{ii},\ t = \frac{z_i}{w_*}\tau,\ \overline{w^2} = u_*^2\overline{w^2_{ii}},\ \overline{u^2} = u_*^2\overline{u_{ii}^2},\ \overline{w\theta} = Q\overline{w\theta}_{ii},\ \overline{\theta^2} = (\frac{Q}{u_*})^2\overline{\theta_{ii}^2},\\
&p = u_*^2p_{ii},
\end{split}\label{inner_inner_variables}
\end{align}
Substituting the inner-inner variables given in (\ref{inner_inner_variables}) into equations (\ref{eq_shear_uw}) to (\ref{eq_t}) we obtain  
\begin{equation}
\frac{u_*^2w_*}{z_i} \frac{\partial{\overline{uw}_{ii}}}{\partial{\tau}} + u_*^2\overline{w^2_{ii}}\frac{U_m}{h_0}\frac{\partial{U_{ii}}}{\partial{z_{ii}}} - \frac{g}{T}Q\overline{u\theta}_{ii} + \frac{u_*^3}{h_0}\Big(\frac{\partial{\overline{uw^2}_{ii}}}{\partial{z_{ii}}} + \Big(\overline{w\frac{\partial{p}}{\partial{x}}}\Big)_{ii} + \Big(\overline{u\frac{\partial{p}}{\partial{z}}}\Big)_{ii}\Big) = 0,
\end{equation}
\begin{equation}
fU_gh_0\frac{w_*}{z_i} \frac{\partial{\overline{vw}_{ii}}}{\partial{\tau}} + u_*^2\overline{w^2_{ii}}\frac{V_g}{h_0}\frac{\partial{V_{ii}}}{\partial{z_{ii}}} - \frac{g}{T}Q\overline{v\theta}_{ii} + \frac{u_*}{h_0}fU_gh_0\Big(\frac{\partial{\overline{vw^2}_{ii}}}{\partial{z_{ii}}} + \Big(\overline{w\frac{\partial{p}}{\partial{y}}}\Big)_{ii} + \Big(\overline{v\frac{\partial{p}}{\partial{z}}}\Big)_{ii}\Big) = 0,
\end{equation}
\begin{equation}
\frac{Qw_*}{z_i} \frac{\partial{\overline{w\theta}_{ii}}}{\partial{\tau}} + u_*^2\overline{w^2_{ii}}\frac{\Theta_m}{h_0}\frac{\partial{\Theta_{ii}}}{\partial{z_{ii}}} + \frac{g}{T}(\frac{Q}{u_*})^2\overline{\theta_{ii}^2} + \frac{Qu_*}{h_0}\Big(\frac{\partial{\overline{w^2\theta}_{ii}}}{\partial{z_{ii}}} + \Big(\overline{\theta\frac{\partial{p}}{\partial{z}}}\Big)_{ii}\Big) = 0,
\end{equation}
\begin{equation}
\frac{u_*^2}{h_0} \frac{\partial{\overline{uw}_{ii}}}{\partial{z_{ii}}} = f(V- V_g) + \frac{d\tau_{rx}}{dz},
\end{equation}
\begin{equation}
fU_g \frac{\partial{\overline{vw}_{ii}}}{\partial{z_{ii}}} = f(U_g - U) + \frac{d\tau_{ry}}{dz},
\end{equation}
\begin{equation}
\frac{\Theta_m}{z_i/w_*} \frac{\partial{\Theta_{ii}}}{\partial{\tau}} + \frac{Q}{h_0} \frac{\partial{\overline{w\theta_{ii}}}}{\partial{z_{ii}}} = 0,
\end{equation}
\begin{equation}
\frac{1}{2} \frac{u_*^2}{z_i/w_*} \frac{\partial{\overline{w_{ii}^2}}}{\partial{\tau}} = -\frac{1}{2} \frac{u_*^3}{h_0} \frac{\partial{\overline{w_{ii}^3}}}{\partial{z_{ii}}}
+ \Big(\overline{p\frac{\partial{w}}{\partial{z}}}\Big)_{ii} \frac{u_*^3}{h_0} - \Big(\overline{\frac{\partial{pw}}{\partial{z}}}\Big)_{ii} \frac{u_*^3}{h_0}
+ \frac{g}{T}Q\overline{w\theta}_{ii} - \varepsilon_3
\end{equation}
\begin{equation}
\frac{1}{2} \frac{u_*^2}{z_i/w_*} \frac{\partial{\overline{u_{ii}^2}}}{\partial{\tau}} = -u_*^2 \overline{uw}_{ii} \frac{u_*}{h_0} \frac{\partial{U_{ii}}}{\partial{z_{ii}}}
-\frac{1}{2} \frac{u_*^3}{h_0} \frac{\partial{\overline{wu_{ii}^2}}}{\partial{z_{ii}}} + \Big(\overline{p\frac{\partial{u}}{\partial{x}}}\Big)_{ii} \frac{u_*^3}{h_0} - \varepsilon_1
\end{equation}
\begin{equation}
\frac{1}{2} (\frac{Q}{u_*})^2 \frac{w_*}{z_i} \frac{\partial{\overline{\theta_{ii}^2}}}{\partial{\tau}} = -Q \overline{w\theta}_{ii} \frac{Q}{u_*h_0} \frac{\partial{\Theta_{ii}}}{\partial{z_{ii}}}
-\frac{1}{2} (\frac{Q}{u_*})^2 \frac{u_*}{h_0} \frac{\partial{\overline{w\theta_{ii}^2}}}{\partial{z_{ii}}} - \varepsilon_{\theta}
\end{equation}
\\
Using $u_*^2U_m/h_0$,  $Qu_*/h_0$, $u_*^2/h_0$, $\Theta_m u_*^2/h_0$,  $u_*^3/h_0$ and $Q^2/u_*h_0$ to non-dimensionalise the shear-stress and vertical flux budgets,
as well as to the mean momentum, mean potential temperature and variance equations, leads to the corresponding governing equations for the inner–inner layer
\begin{equation}
  \frac{h_0}{z_i}\frac{w_*}{U_m}\frac{\partial{\overline{uw}_{ii}}}{\partial{\tau}} + \overline{w^2_{ii}}\frac{\partial{U_{ii}}}{\partial{z_{ii}}} - \frac{(g/T)Qh_0}{u_*^2U_m}\overline{u\theta}_{ii} + \epsilon_1' \Big(\frac{\partial{\overline{uw^2}_{ii}}}{\partial{z_{ii}}} + \Big(\overline{w\frac{\partial{p}}{\partial{x}}}\Big)_{ii} + \Big(\overline{u\frac{\partial{p}}{\partial{z}}}\Big)_{ii}\Big) = 0,
  \label{eq:ii_uw}
\end{equation}
\begin{equation}
\frac{h_0}{z_i}\frac{fh_0w_*}{u_*^2}\frac{U_g}{U_m}\frac{\partial{\overline{vw}_{ii}}}{\partial{\tau}} + \overline{w^2_{ii}}\frac{V_g}{U_m}\frac{\partial{V_{ii}}}{\partial{z_{ii}}} - \frac{(g/T)Qh_0}{u_*^2U_m}\overline{v\theta}_{ii} + \epsilon_2'' \frac{V_g}{U_m}\Big(\frac{\partial{\overline{vw^2}_{ii}}}{\partial{z_{ii}}} + \Big(\overline{w\frac{\partial{p}}{\partial{y}}}\Big)_{ii} + \Big(\overline{v\frac{\partial{p}}{\partial{z}}}\Big)_{ii}\Big) = 0,
\end{equation}
\begin{equation}
\epsilon_3' \epsilon_3'' \epsilon_{\theta}' \frac{\partial{\overline{w\theta}_{ii}}}{\partial{\tau}} + \overline{w^2_{ii}}\frac{\partial{\Theta_{ii}}}{\partial{z_{ii}}} + \epsilon_3'' \epsilon_{\theta}' \overline{\theta_{ii}^2} + \epsilon_{\theta}' \Big(\frac{\partial{\overline{w^2\theta}_{ii}}}{\partial{z_{ii}}} + \Big(\overline{\theta\frac{\partial{p}}{\partial{z}}}\Big)_{ii}\Big) = 0,
\end{equation}
\begin{equation}
\frac{\partial{\overline{uw}_{ii}}}{\partial{z_{ii}}} = \epsilon_5'' - \epsilon_5'' \epsilon_2'' V_{ii,2} + \frac{d\tau_{rxii}}{dz_{ii}}
\end{equation}
\begin{equation}
\frac{\partial{\overline{vw}_{ii}}}{\partial{z_{ii}}} = 1 - \epsilon_4'' U_{ii,2} + \frac{d\tau_{ryii}}{dz_{ii}}
\end{equation}
\begin{equation}
\epsilon_3' \epsilon_3'' \frac{\partial{\Theta_{ii}}}{\partial{\tau}} + \epsilon_{\theta}' \frac{\partial{\overline{w\theta_{ii}}}}{\partial{z_{ii}}} = 0,
\end{equation}
\begin{equation}
\frac{1}{2} \epsilon_3' \epsilon_3'' \frac{\partial{\overline{w_{ii}^2}}}{\partial{\tau}} = -\frac{1}{2} \frac{\partial{\overline{w_{ii}^3}}}{\partial{z_{ii}}}
+ \Big(\overline{p\frac{\partial{w}}{\partial{z}}}\Big)_{ii} - \Big( \overline{\frac{\partial{pw}}{\partial{z}}} \Big)_{ii} + \epsilon_3'' \overline{w\theta}_{ii} - \varepsilon_{3ii}
\label{eq:ii_w2}
\end{equation}
\begin{equation}
\frac{1}{2} \epsilon_3' \epsilon_3'' \frac{\partial{\overline{u_{ii}^2}}}{\partial{\tau}} = -\overline{uw}_{ii} \frac{\partial{U_{ii}}}{\partial{z_{ii}}}
- \frac{1}{2} \frac{\partial{\overline{wu_{ii}^2}}}{\partial{z_{ii}}} + \Big(\overline{p\frac{\partial{u}}{\partial{x}}}\Big)_{ii} - \varepsilon_{1ii}
\label{eq:ii_u2}
\end{equation}
\begin{equation}
\frac{1}{2} \epsilon_3' \epsilon_3'' \frac{\partial{\overline{\theta_{ii}^2}}}{\partial{\tau}} = -\overline{w\theta}_{ii} \frac{\partial{\Theta_{ii}}}{\partial{z_{ii}}}
- \frac{1}{2} \frac{\partial{\overline{w\theta_{ii}^2}}}{\partial{z_{ii}}} - \varepsilon_{\theta ii}
\end{equation}
where
\begin{align}
\epsilon_1' = \frac{u_*}{U_m},\ 
\epsilon_2'' = \frac{fh_0}{u_*}\frac{U_g}{V_g},\ 
\epsilon_3'' = \frac{h_0}{L},\\
\epsilon_4'' = \frac{u_*}{U_g},\ 
\epsilon_5'' = \frac{h_0}{z_i},\ 
\epsilon_{\theta}' = \frac{Q}{u_* \Theta_m}
\end{align}
Here $\epsilon_2''$ and $\epsilon_4''$ quantify the effects of the Coriolis force in the $U$- and $V$-components of the mean velocity.
$\epsilon_3''$ quantifies the effects of the buoyancy production of $\overline {w^2}$. $\epsilon_5''$ quantifies the effects of variations of the
shear stress with height.
$\epsilon_\theta''$ represents the relative scale of the mean temperature.\\

The inner-inner expansions are
\begin{equation}
\overline{w_{ii}^2}(z_{ii}) = \overline{w_{ii}^2}_{,1}(z_{ii}) + \epsilon_3'' \overline{w^2_{ii,2}(z_{ii})} + O(\epsilon_3''^2),
\end{equation}
\begin{equation}
U_{ii} = U_{ii,1} + \epsilon_1' U_{ii,2} + \epsilon_1' \epsilon_3'' U_{ii,31} + \epsilon_1' \epsilon_5'' U_{ii,32} + \cdot\cdot\cdot
\end{equation}
\begin{equation}
\frac{\partial{\overline{uw^2_{ii}}}}{\partial{z_{ii}}} = \frac{\partial{\overline{uw^2_{ii,1}}}}{\partial{z_{ii}}} + \epsilon_3'' \frac{\partial{\overline{uw^2_{ii,2}}}}{\partial{z_{ii}}}
+ \epsilon_5'' \epsilon_2'' \frac{\partial{\overline{uw^2_{ii,3}}}}{\partial{z_{ii}}} + \cdot\cdot\cdot
\end{equation}
\begin{equation}
V_{ii} = V_{ii,1} + \epsilon_2'' V_{ii,2} + \epsilon_2'' \epsilon_3'' V_{ii,31} + \epsilon_2'' \epsilon_4'' V_{ii,32} + \cdot\cdot\cdot
\end{equation}
\begin{equation}
\frac{\partial{\overline{vw^2_{ii}}}}{\partial{z_{ii}}} = \frac{\partial{\overline{vw^2_{ii,1}}}}{\partial{z_{ii}}} + \epsilon_3'' \frac{\partial{\overline{vw^2_{ii,21}}}}{\partial{z_{ii}}}
+ \epsilon_4'' \frac{\partial{\overline{vw^2_{ii,22}}}}{\partial{z_{ii}}} + \cdot\cdot\cdot
\end{equation}
\begin{equation}
\Theta_{ii} = \Theta_{ii,1} + \epsilon_{\theta}' \Theta_{ii,2} + \epsilon_{\theta}' \epsilon_3'' \Theta_{ii,3} + \epsilon_{\theta}' \epsilon_3' \epsilon_3'' \Theta_{ii,4} + \cdot\cdot\cdot
\end{equation}
Substitute the inner-inner expansions back to equations (\ref{eq:ii_uw}), (\ref{eq:ii_w2}) and (\ref{eq:ii_u2}) we obtain 
\begin{align}
\begin{split}
(\overline{w_{ii,1}^2} + \epsilon_3'' \overline{w^2_{ii,2}} + \cdot\cdot\cdot) \Big(\frac{\partial{U_{ii,1}}}{\partial{z_{ii}}} + \epsilon_1' \frac{\partial{U_{ii,2}}}{\partial{z_{ii}}} + \epsilon_1' \epsilon_3'' \frac{\partial{U_{ii,31}}}{\partial{z_{ii}}}  + \epsilon_1' \epsilon_5'' \frac{\partial{U_{ii,32}}}{\partial{z_{ii}}} + \cdot\cdot\cdot\Big)\\
  + \epsilon_1'\Big(\Big(\overline{w\frac{\partial{{p}}}{\partial{x}}}\Big)_{ii,1} + \epsilon_3'\Big(\overline{w\frac{\partial{{p}}}{\partial{x}}}\Big)_{ii,2} + \epsilon_5' \epsilon_2'\Big(\overline{w\frac{\partial{{p}}}{\partial{x}}}\Big)_{ii,3}\Big) + \cdot\cdot\cdot = 0.
\end{split}
\end{align}

\begin{align}
\begin{split}
\frac{1}{2}\epsilon_3'\epsilon_3''\Bigg(
\frac{\partial \overline{w^2_{ii,1}}}{\partial \tau}
+ \epsilon_3'' \frac{\partial \overline{w^2_{ii,2}}}{\partial \tau} + \cdot\cdot\cdot
\Bigg)
= -\frac{1}{2}\Bigg(
\frac{\partial \overline{w^3_{ii,1}}}{\partial z_{ii}}
+ \epsilon_3'' \frac{\partial \overline{w^3_{ii,2}}}{\partial z_{ii}} + \cdot\cdot\cdot
\Bigg)\\ + \cdot\cdot\cdot + \Big(\overline{w\theta}_{ii,1} + \epsilon_\theta' \overline{w\theta}_{ii,2} + \cdot\cdot\cdot\Big) - \varepsilon_{3ii}
\end{split}
\end{align}

\begin{align}
\begin{split}
\frac{1}{2}\epsilon_3'\epsilon_3''\frac{\partial \overline{u^2_{ii}}}{\partial \tau}
&= - \Big(\overline{uw}_{ii,1} + \epsilon_3''\overline{uw}_{ii,2} + \epsilon_5'' \epsilon_2''\overline{uw}_{ii,3} + \cdot\cdot\cdot\Big)\Big(\frac{\partial{U_{ii,1}}}{\partial{z_{ii}}} + \epsilon_1' \frac{\partial{U_{ii,2}}}{\partial{z_{ii}}}\\
& + \epsilon_1'\epsilon_3'' \frac{\partial{U_{ii,31}}}{\partial{z_{ii}}}
 + \epsilon_1'\epsilon_5'' \frac{\partial{U_{ii,32}}}{\partial{z_{ii}}} + \cdot\cdot\cdot\Big) - \frac{1}{2}\Big(
\frac{\partial \overline{uw^2}_{ii,1}}{\partial z_{ii}}
+ \epsilon_3'' \frac{\partial \overline{uw^2}_{ii,21}}{\partial z_{ii}} \\
& + \epsilon_5'' \epsilon_2'' \frac{\partial \overline{uw^2}_{ii,22}}{\partial z_{ii}} + \cdot\cdot\cdot
\Big) + \cdot\cdot\cdot - \varepsilon_{1ii}
\end{split}
\end{align}
\\
The buoyancy production of $\overline{w^2_{ii,1}}$ is of higher order, but becomes leading order in the inner-outer layer. Therefore, it induces the second-order terms in
the inner-inner expansions.

\subsection{Matching between outer and inner-outer layers}

The asymptotic expansions for each layer are valid as long as the location $z$ is away from the other layers. For example, the outer expansions
are valid for $-z/L \gg 1$ while inner-outer expansions are valid for $z/z_i \ll 1$ and $z/h_0 \gg 1$.  As a result, for $-L \ll z \ll z_i$
both the outer and inner-outer expansions are valid. We can asymptotically match the expansions in this so-called matching or overlapping layer to obtain the scaling of the terms in the 
expansions. Therefore, asymptotic matching effectively provides additional equations to close the equation set, allowing us to obtain the functional form of the profile in the matching layer,
and avoiding the need for a closure model.

We now asymptotically match the third-order outer expansion of the mean velocity with the second-order inner-outer expansion,
\begin{equation}
U = U_m ( U_{o,1} + \epsilon_1 U_{o,2} + \epsilon_1 \epsilon_3 U_{o,31} + \epsilon_1\epsilon_3^2 U_{o,4} + \cdot\cdot\cdot ),
\end{equation}
\begin{equation}
U = U_m ( U_{io,1} + \epsilon_1' U_{io,2} + \epsilon_1' \epsilon_3' U_{io,31} +  \epsilon_1' \epsilon_5' U_{io,32} + \cdot\cdot\cdot).
\end{equation}
Here we include the $\epsilon_1' \epsilon_5'$ term which is smaller than $\epsilon_1' \epsilon_3'$, but is
larger than the next-order  term $\epsilon_1' \epsilon_3'^{2}$. In general, matching is performed by taking
intermediate limits $\epsilon_i\rightarrow 0$ and $z_o\rightarrow 0$ for the outer expansions and
$\epsilon'_i\rightarrow 0$ and $z_{io}\rightarrow \infty$ for the inner-outer expansions
(\citealt{BO1978}). However, matching between the terms of the same order alone is
often insufficient. According to van Dyke's matching principle (\citealt{vanDyke1975}), the same number of terms in
the inner expansion of the outer expansion
and the outer expansion of the inner expansion need to be matched. Therefore, matching across the orders is generally necessary. 

Matching the expansions of the same orders results in
\begin{equation}
U_{o,1} = U_{io,1},\ \ 
\epsilon_1 U_{o,2} = \epsilon_1' U_{io,2} \footnote{Strictly speaking, this equation should be
  $\lim\limits_{\epsilon_1,\,z_o \to 0} \epsilon_1 U_{o,2} = \lim\limits_{\epsilon_1' \to 0,\ z_{io} \to \infty} \epsilon_1' U_{io,2}$. To simplify the notations, we omit the limits.
  The other equations are written in a similar manner.}, \\ 
\epsilon_1 \epsilon_3 U_{o,31} = \epsilon_5' \epsilon_1' U_{io,32},\ \ 
\epsilon_1 \epsilon_3 U_{o,31} = \epsilon_1' \epsilon_3' U_{io,31},
\end{equation}
leading to
\begin{equation}
U_{o,1} = U_{io,1} = 1,\ 
\end{equation}
\begin{equation}
U_{o,21} = A (\frac{z}{z_i})^{-1/3},\ 
U_{io,21} = A (-\frac{z}{L})^{-1/3},\ 
\end{equation}
\begin{equation}
U_{o,311} = B (\frac{z}{z_i})^{-2/3},\ 
U_{io,321} = B (-\frac{z}{L})^{-2/3},\ 
\end{equation}
\begin{equation}
U_{o,312} = J (\frac{z}{z_i})^{-1},\ 
U_{io,311} = J (-\frac{z}{L})^{-1}. 
\end{equation}

Matching the expansions of different orders (cross-order matching) gives
\begin{equation}
\epsilon_1 U_{o,2} = \epsilon_1' \epsilon_3' U_{io,31},\ 
\epsilon_1 \epsilon_3 U_{o,31} = \epsilon_1' U_{io,2},\
\epsilon_1 U_{o,2} =  \epsilon_1' \epsilon_5' U_{io,32}.\
\end{equation}
\begin{equation}
\epsilon_1 \epsilon_3^2 U_{o,41} = \epsilon_1' U_{io,2},\
\epsilon_1 \epsilon_3^2 U_{o,42} =  \epsilon_1' \epsilon_5' U_{io,32},\
\epsilon_1 \epsilon_3^2 U_{o,43} = \epsilon_1' \epsilon_3' U_{io,31},\
\end{equation}
The results of the matching process (details given in Appendix~\ref{app:cross}) are
\begin{equation}
U_{o,22} = D(\frac{z}{z_i})^{1/3},\ 
U_{io,312} = D(-\frac{z}{L})^{1/3},\ 
\end{equation}
\begin{equation}
U_{o,313} = E(\frac{z}{z_i})^{-5/3},\ 
U_{io,22} = E(-\frac{z}{L})^{-5/3},\ 
\end{equation}
\begin{equation}
U_{o,23} = F(\frac{z}{z_i})^{2/3},\
U_{io,322} = F(-\frac{z}{L})^{2/3},\
\end{equation}
\begin{equation}
U_{o,41} = G(\frac{z}{z_i})^{-3},\
U_{io,23} = G(-\frac{z}{L})^{-3},\
\end{equation}
\begin{equation}
U_{o,42} = H(\frac{z}{z_i})^{-2},\
U_{io,323} = H(-\frac{z}{L})^{-2},\
\end{equation}
\begin{equation}
U_{o,43} = I(\frac{z}{z_i})^{-7/3},\
U_{io,313} = I(-\frac{z}{L})^{-7/3},\
\end{equation}

Inserting the matching results into equations (\ref{eq:Uo}) and (\ref{eq:Uio}), we obtain the higher-order asymptotic expansions for the local-free-convection layer
in terms of the outer and inner-outer variables respectively
\begin{equation}
  \begin{split}
  U_o &= U_{o,1} + \epsilon_1 U_{o,21} + \epsilon_1 U_{o,22} + \epsilon_1 U_{o,23} + \epsilon_1 \epsilon_3 U_{o,311} + \epsilon_1 \epsilon_3 U_{o,312} + \epsilon_1 \epsilon_3 U_{o,313}\\
&\quad +\epsilon_1 \epsilon_3^2 U_{o,41} + \epsilon_1 \epsilon_3^2 U_{o,42} + \epsilon_1 \epsilon_3^2 U_{o,43} + \cdot\cdot\cdot\\
  &\quad= 1 + \epsilon_1 A(\frac{z}{z_i})^{-1/3} + \epsilon_1 D(\frac{z}{z_i})^{1/3} + \epsilon_1 F(\frac{z}{z_i})^{2/3} + \epsilon_1 \epsilon_3 B (\frac{z}{z_i})^{-2/3}+ \epsilon_1 \epsilon_3 J(\frac{z}{z_i})^{-1}\\
  &\quad+ \epsilon_1 \epsilon_3 E(\frac{z}{z_i})^{-5/3} + \epsilon_1 \epsilon_3^2 G(\frac{z}{z_i})^{-3} + \epsilon_1 \epsilon_3^2 H(\frac{z}{z_i})^{-2} + \epsilon_1 \epsilon_3^2 I(\frac{z}{z_i})^{-7/3} + \cdot\cdot\cdot\\   
\end{split}
  \label{eq:match_o}
\end{equation}

\begin{equation}
  \begin{split}
  U_{io} &= U_{io,1} + \epsilon_1' U_{io,21}+ \epsilon_1' U_{io,22} + \epsilon_1' U_{io,23}+ \epsilon_1' \epsilon_3' U_{io,311}+ \epsilon_1' \epsilon_3' U_{io,312}+ \epsilon_1' \epsilon_3' U_{io,313}\\
  &\quad + \epsilon_5' \epsilon_1' U_{io,321} + \epsilon_5' \epsilon_1' U_{io,322} + \epsilon_5' \epsilon_1' U_{io,323} + \cdot\cdot\cdot\\
  &\quad= 1 + \epsilon_1' A(-\frac{z}{L})^{-1/3} + \epsilon_1' E(-\frac{z}{L})^{-5/3} + \epsilon_1' G(-\frac{z}{L})^{-3} + \epsilon_1' \epsilon_3' J (-\frac{z}{L})^{-1}  + \epsilon_1' \epsilon_3' D (-\frac{z}{L})^{1/3}\\
 &\quad + \epsilon_1' \epsilon_3' I (-\frac{z}{L})^{-7/3} + \epsilon_5' \epsilon_1'B (-\frac{z}{L})^{-2/3} + \epsilon_5' \epsilon_1'F(-\frac{z}{L})^{2/3} + \epsilon_5' \epsilon_1' H(-\frac{z}{L})^{-2} + \cdot\cdot\cdot\\
\end{split}
\label{eq:match_io}
\end{equation}

\vspace{10pt}

\subsection{Matching between inner-outer and inner-inner layer}

Matching mean velocity between the inner-outer and inner-inner layer is performed by taking the intermediate limits $\epsilon'_i\rightarrow 0$ and $ z_{io}\rightarrow 0$ for
the inner-outer expansions, and $\epsilon'_i\rightarrow 0$ and $ z_{ii}\rightarrow \infty$ for the inner-outer expansions. The expansions are
\begin{equation}
U_{io} = U_{io,1} + \epsilon_1' U_{io,2} + \cdot\cdot\cdot
\end{equation}
\begin{equation}
U_{ii} = U_{ii,1} + \epsilon_1' U_{ii,2} + \epsilon_1' \epsilon_3'' U_{ii,32} + \epsilon_1' \epsilon_3''^2 U_{ii,41} + \cdot\cdot\cdot
\end{equation}
Matching the expansion terms of the same order results in the log law 
\begin{equation}
U_{io,1} + \epsilon_1' U_{io,2} = \epsilon_1' U_{ii,2},\ 
\end{equation}
\begin{equation}
U_{io,21} = A' \frac{1}{\kappa} \ln(-\frac{z}{L}) + C,\ 
U_{ii,2} = A' \frac{1}{\kappa} \ln(\frac{z}{h_0}), 
\end{equation}
Cross-order matching (details are in Appendix~\ref{app:cross}) gives
\begin{equation}
\epsilon_1' U_{io,2} = \epsilon_1' \epsilon_3'' U_{ii,32},\
\epsilon_1' U_{io,2} = \epsilon_1' \epsilon_3''^2 U_{ii,41},\
\end{equation}
\begin{equation}
U_{io,22} = C'(-\frac{z}{L}),\ 
U_{ii,32} = C'\frac{z}{h_0}, 
\end{equation}
\begin{equation}
U_{io,23} = E'(-\frac{z}{L})^2,\ 
U_{ii,41} = E'(\frac{z}{h_0})^2, 
\end{equation}
The matching results in terms of the inner-outer and inner variables are
\begin{align}\label{eq:inner_matching_io}
\begin{split}
  U_{io} &= U_{io,1} + \epsilon_1' U_{io,21} + \epsilon_1' U_{io,22}  + \epsilon_1' U_{io,23} + \cdot\cdot\cdot 
  \\
&= 1 + \epsilon_1' A' \frac{1}{\kappa} \ln(-\frac{z}{L}) + \epsilon_1' C + \epsilon_1' C' (-\frac{z}{L}) + \epsilon_1' E' (-\frac{z}{L})^2 + \cdot\cdot\cdot,
\end{split}
\end{align}
\begin{align}\label{eq:inner_matching_ii}
\begin{split}
  U_{ii} &= U_{ii,1} + \epsilon_1' U_{ii,2} + \epsilon_1' \epsilon_3'' U_{ii,32} + \epsilon_1' \epsilon_3''^2 U_{ii,41} + \cdot\cdot\cdot 
  \\
&=  \epsilon_1' A' \frac{1}{\kappa} \ln(\frac{z}{h_0}) + \epsilon_1' \epsilon_3'' C' \frac{z}{h_0} + \epsilon_1' \epsilon_3''^2 E' (\frac{z}{h_0})^{2} + \cdot\cdot\cdot,
\end{split}
\end{align}

Equating the RHS of equations (\ref{eq:inner_matching_io}) and (\ref{eq:inner_matching_ii}) the convective logarithmic friction law is obtained
\begin{equation}
  \frac{U_m}{u_*} + C = \frac{1}{\kappa}\ln(-\frac{L}{h_0}),
\end{equation}
This friction law, while identical to the leading-order friction law derived by \cite{TD19b}, is valid up to the
second order (likely is valid to all orders as we do not foresee any higher-order logarithmic expansion terms).
Therefore, there are no other parameters in the friction law.
This situation is in contrast to channel and pipe flows with smooth walls, where the logarithmic frictional law is
a leading-order approximation (\citealt{Tennekes68,Afzal1976}).

\section{Determination of expansion coefficients using field data}

\subsection{Field data}

The asymptotic expansions provide the functional forms of the higher-order profile in the matching layers, However, they contain non-dimensional universal expansion
coefficients, which need to be determined using field measurement data.
To obtain the coefficients and to validate the expansions, a field campaign code-named Multi-point Monin–Obukhov Similarity Horizontal Array Turbulence Study (M$^2$HATS)
was conducted from July to September
2023 in Tonopah, Nevada, the United States. A comprehensive suite of in situ and remote sensing instruments was deployed to characterize the convective
boundary layer under approximately horizontally homogeneous conditions (\citealt{Tong2026a}). The dataset used for obtaining the mean profiles
includes quality-controlled surface meteorology and flux measurements from towers including 3D sonic anemometers, and data from a Doppler lidar system (WindCube 200S).
The anemometer were mounted at multiple heights (0.62, 1.17, 2.11, 3.02, 4.20, 6.89, 15.45,  28.55 m) and had a native
sampling rate of 60 samples/s. The lidar   performed continuous plan–position indicator (PPI) scans
at a 35.3$^\circ$ elevation angle, with a azimuth coverage from 2$^\circ$ to 360$^\circ$, an angular resolution of 2.5$^\circ$, and a typical scan rate of 1 revolution per minute.
These measurements provide coverage from 0.62 m 
to over 3 km above ground level, 
ensuring accurate determination of stability parameters
and mixed-layer variables such as $u_*$, $L$, and $z_i$. This  dataset covers a range of $z_i/L$ and $h_0/L$ values, enabling validation of the leading-order velocity profile
in the surface layer (the log-law and the local-free-convection scaling) as well as determination of the non-dimensional coefficients
in the higher-order asymptotic expansions 
that account for finite scale separations ($z_i/L$ and $h_0/L$), thereby providing improved characterization of the mean velocity profile
in the CBL.


\vspace{0.2in}
\begin{figure}
  \centering
  \includegraphics[width=1\linewidth]{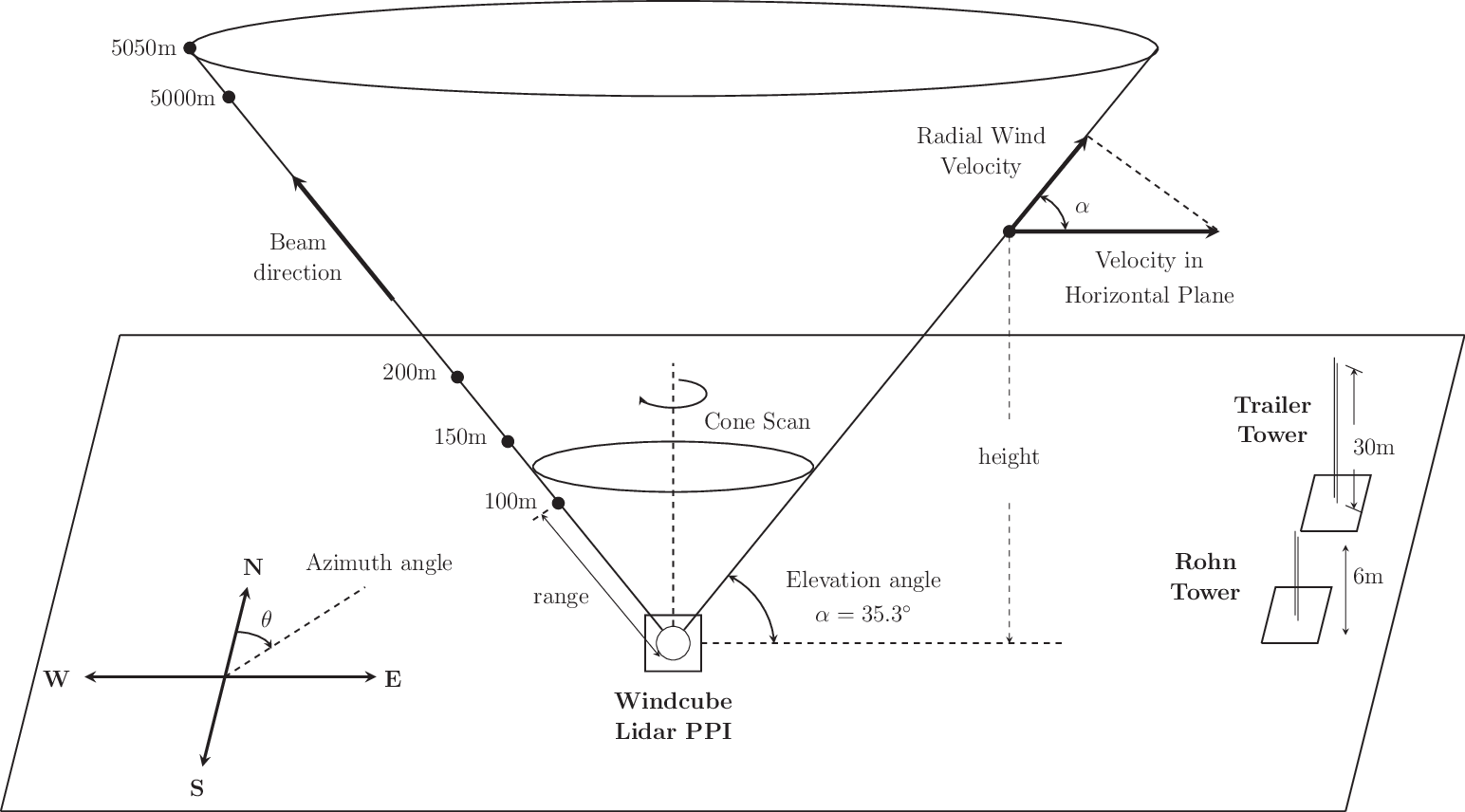}
  \caption{Schematic of the measurement set-up for obtaining the mean profile. Two met towers, 6 m and 32 m in height respectively, provided the mean velocity, the 
    mean shear stress, flux, and gradients near the surface.
    A Doppler
 lidar performing Plan Position Indicator (PPI or cone) scan, was used to measure the mean horizontal velocity profiles throughout the boundary layer. }
  \label{fig:Lidar}
\end{figure}

During the field campaign, daily catalogs (07:00–21:00 local time) were compiled, including: vertical lidar $w$-wind profiles, sky photographs every two hours,
virtual potential temperature and mixing ratio profiles from sondes, potential temperature, humidity, and backscatter measurements
from MicroPulse Differential Absorption Lidar (MPD), 10-minute statistics from sonic anemometers, raw $u$, $v$, $w$, $T$ signals from the sonic anemometers.
From these data, 91 stationary periods were manually selected for the flux towers. The boundary-layer height $z_i$ was
determined using the MPD, lidar, and sonde measurements.
Among the periods, 14  were selected for the barotropic cases (\citealt{Tong2026a})

\subsection{Procedure for the local-free-convection layer}
  
The data analysis  for the local-free-convection layer is conducted using a multi-step procedure to accurately characterising
the mean velocity profile and to obtain the expansion coefficients. The procedure includes (i) selecting suitable periods and extracting velocity data
from the Doppler lidar signals, (ii) identifying the suitable height range for the leading-order scaling, (iii) iteratively
estimating the expansion coefficients using a combination of linear regression, (iv) stabilizing the coefficients using
ridge regression (\citealt{hoerl1970ridge}), and (v) quantifying the uncertainties of the coefficients via bootstrap resampling. The following sections
describe each of these steps in detail.

\subsubsection{Determination of leading-order matching region}

To identify the height range in which the leading-order scaling of the mean velocity (defect) profile, $(z/L)^{-1/3}$, is approximately valid,
we perform a systematic search for the lower (in terms of $-z/L$) and upper (in terms of $z/z_i$) bounds of each profile. 
To directly identify the height range in which the leading-order scaling of the mean velocity (defect) profile, $(z/L)^{-1/3}$, is approximately valid,
 one needs the mixed-layer velocity scale $U_m$, which is not available at this point of the data analysis. To circumvent the need for $U_m$, for each
profile we average the velocity defect over the scaling layer
\begin{equation}
\left( \overline{\frac{U}{u_*}}-\frac{U_m}{u_*}\right)
= a  \overline{(z/L)^{-1/3}}  ,
\end{equation}
where the coefficient $a$ has the same value for all the profiles. Subtracting this equation from the velocity defect law, we have
\begin{equation}\label{heightID}
\left( \overline{\frac{U}{u_*}}-\frac{U}{u_*} \right)
= a \left[  \overline{(z/L)^{-1/3}} -(z/L)^{-1/3} \right].
\end{equation}
If the data approximately satisfies equation (\ref{heightID}),  a height range for the $(z/L)^{-1/3}$ scaling has been identified 
without using $U_m/u_*$. Since $a$ is the same for all the profiles which have different $U_m/u_*$ values, we
can combine all the data to evaluate the accuracy of the linear relationship.
We then perform a least-squares linear regression, and use the coefficient of determination (\(R^2\)) to quantify the goodness of fit.

\begin{figure}
  \centering
  \includegraphics[width=1\linewidth]{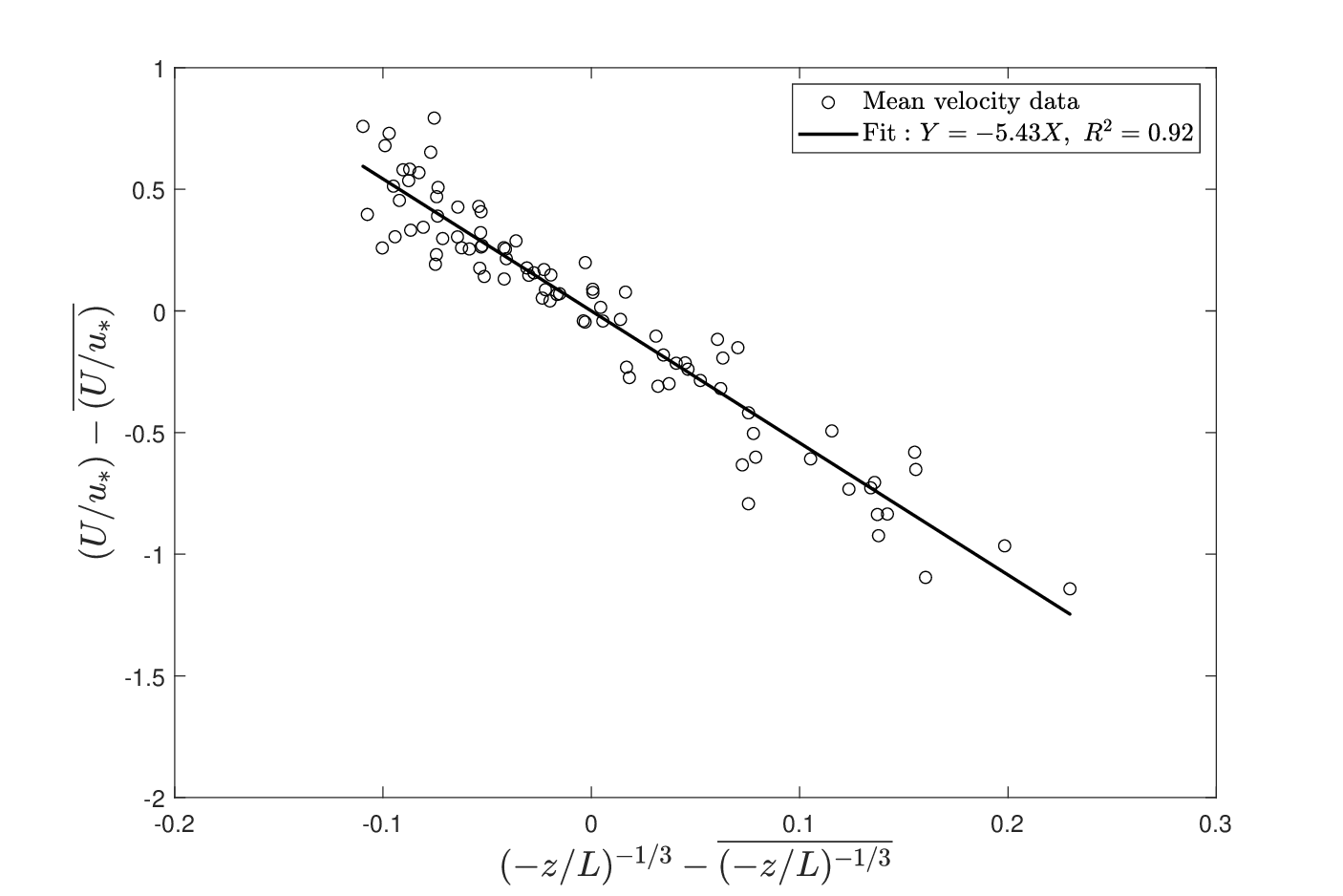}
  \caption{Selected non-dimensional height range for matching between the outer and inner-outer layers}
  \label{fig:height_range}
\end{figure}

Starting from the range $z=-L$ to $z=0.2 z_i$ and progressively reducing it, the optimal height range is identified as the one yielding the maximum \(R^2\), i.e.,
the range where the data most closely follow the theoretical
leading-order scaling. 
In figure~\ref{fig:height_range}, the selected interval for matching layer of outer and inner-outer layers has shown.

\subsubsection{Iterative procedure for determining expansion coefficients}\label{subsec:lfco}

Here we briefly describe the procedure and provide the details in Appendix~\ref{app:lfco}.
The non-dimensional velocity profile is expressed as an asymptotic expansion in the powers of
$(-z/L)$, including one leading-order term and three higher-order terms.
The resulting expression used for the parameter estimation is 
\begin{equation}
    \frac{U}{u_*} = \frac{U_m}{u_*} + A\left(-\frac{z}{L}\right)^{-1/3}
    + E\left(-\frac{z}{L}\right)^{-5/3}
    + G\left(-\frac{z}{L}\right)^{-3}
    + \epsilon_3' D\left(-\frac{z}{L}\right)^{1/3},
    \label{eq:FCL}
\end{equation}
The coefficients $A$, $E$, $D$, $G$, and the $U_m/u_\ast$ values are determined using an iterative multi-step procedure which it is explained in Appendix~\ref{app:lfco}. 
We do not include the terms containing the coefficients $F$, $H$, $I$ and $J$ (equations (\ref{eq:match_o}) and (\ref{eq:match_io})) in the regression due to the size of the data.
To stabilize the regression coefficients and prevent unrealistically large values of the higher-order coefficients,
we employ ridge regression as a regularization strategy (\citealt{willoughby1979solutions}). The mathematical formulation of the ridge-regularized
regression and implementation details are provided in Appendix~\ref{app:ridge}.
To assess the statistical uncertainty of the fitted coefficients, we apply a bootstrap resampling approach,
which enables estimation of coefficient variability without assuming any particular error distribution.
The bootstrap methodology and its implementation in the present analysis are described in Appendix~\ref{app:Bootstrap}.

\subsection{Procedure for the log layer}

In the surface layer, for $z>-0.1L$ the deviations of measurement data from the logarithmic form
is clearly visible. As a result, direct identification of the log layer and the determination of the
von K\'{a}rm\'{a}n constant is nearly impossible. However, the results obtained from matching of the
inner-outer and inner-inner layers show that, by including the first and second higher-order terms, 
we can still obtain the coefficients in the higher-order expansions  as well as the von K\'{a}rm\'{a}n
constant from the measurements.

The expansions in terms of the inner-outer and inner-inner variables are
\begin{align}
\begin{split}
U_{io} &= 1 + \epsilon_1' \frac{1}{\kappa} \ln(-\frac{z}{L}) + \epsilon_1' C' (-\frac{z}{L}) + \epsilon_1' C'\alpha (-\frac{z}{L})^2 + \epsilon_1' C,
\end{split}
\end{align}
\begin{align}
\begin{split}
U_{ii} &=  \epsilon_1' \frac{1}{\kappa} \ln(\frac{z}{h_0}) + \epsilon_1' \epsilon_3'' C' \frac{z}{h_0} + \epsilon_1' \epsilon_3''^2 C'\alpha (\frac{z}{h_0})^{2},
\end{split}\label{eq:ii_expansion}
\end{align}
The second equation can be written as
\begin{equation}
\frac{U}{u_*}
= \frac{1}{\kappa}\ln\!\left(\frac{z}{h_0}\right)
+ C'\left(\epsilon_3^{\prime\prime}\,\frac{z}{h_0}
+ \alpha \,(\epsilon_3^{\prime\prime})^{2}\!\left(\frac{z}{h_0}\right)^{2}\right),
\qquad \epsilon_3'' = \frac{h_0}{-L}
\end{equation}
which can be further written as
\begin{equation}
  \frac{U}{u_*} = \frac{1}{\kappa}\ln(z) - \frac{1}{\kappa}\ln(h_0) + C'(-\frac{z}{L}) + C'\alpha(-\frac{z}{L})^2,
  \label{eq:log}
\end{equation}
where $C'\alpha = E'$ is a higher-order expansion coefficient.
To determine the expansion coefficients, a linear regression is performed to fit the equation
\begin{equation}
  Y = \frac{1}{\kappa}X_1 - \frac{1}{\kappa}\ln(h_0) + C'X_2 + C'\alpha X_3,
  \label{eq:log_fit}
\end{equation}
where
\begin{equation}
Y = \frac{U}{u_*}, \quad 
X_1 = \ln(z), \quad 
X_2 =(-\frac{z}{L}), \quad
X_3 =(-\frac{z}{L})^2,
\end{equation}
The term $-\dfrac{1}{\kappa}\ln(h_0)$ is the intercept and determines the roughness height.
In this analysis, the regression procedure is applied to the flux tower data. Specifically, data points are considered
for heights ranging from $z = 1\,\mathrm{m}$ up to $z = 1.3|L|$. We  find that the higher-order asymptotic
expansions agree well with the data within this height range.
To quantify the statistical uncertainty associated with the estimated expansion coefficients  
($\kappa$, $C'$, $C'\alpha$, and $h_0$), we again employ the bootstrap resampling approach
described in Appendix~\ref{app:Bootstrap}. 

\section{Expansion coefficients and validation of prediction}

In this section we will discuss the higher-order expansion coefficients obtained from matching
the outer and inner-outer layers, and from matching the inner-outer and inner-inner layers.
These coefficients have been obtained based on the methods discussed in the previous section.
The resulting asymptotic expansions are validated with the data. 

\begin{figure}
  \centering
  \includegraphics[width=1\linewidth]{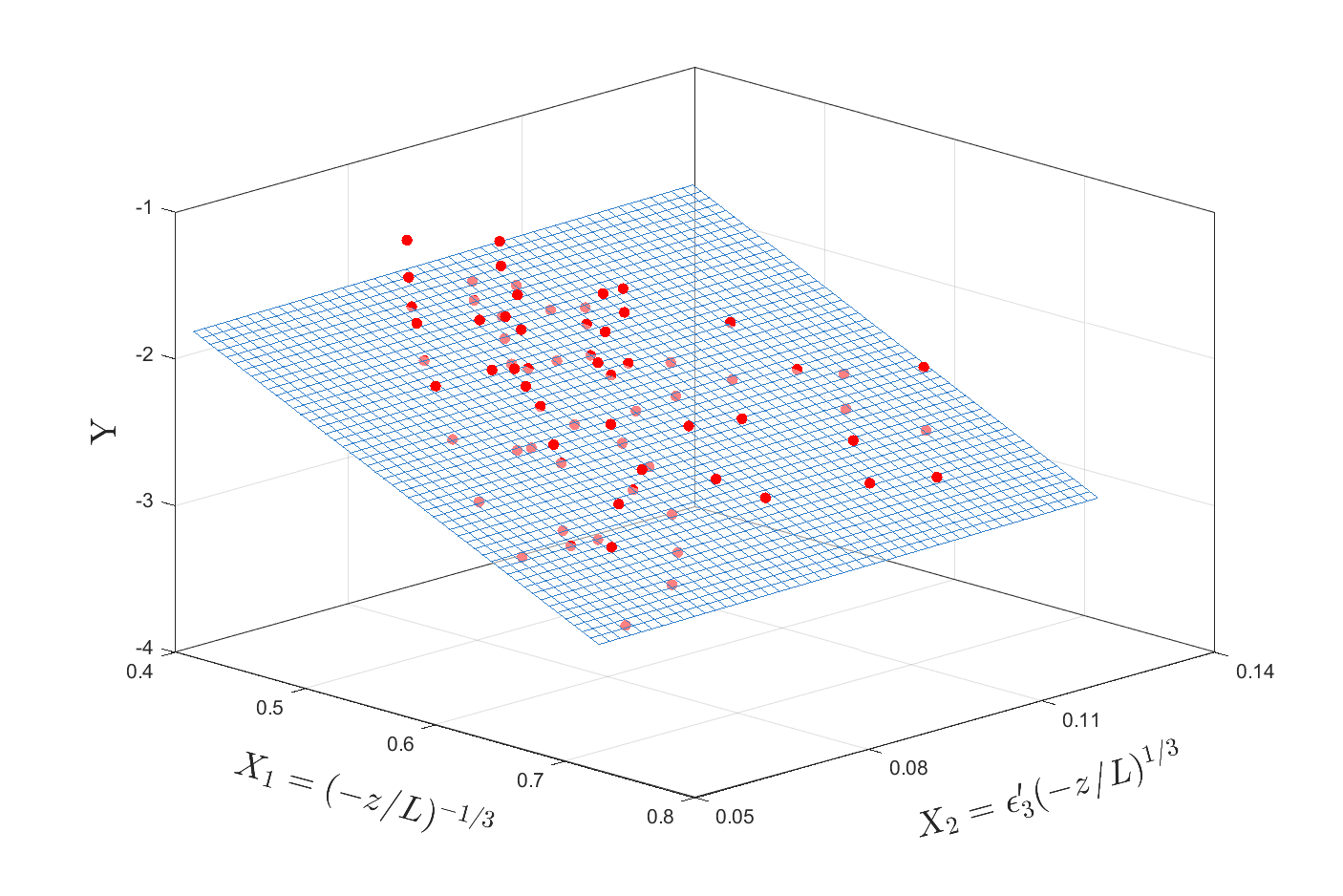}
  \caption{Planar fit of the expansion coefficients to data in the matching region between the outer and inner-outer layers (local-free-convection layer).
    The variable, $Y = \dfrac{U}{u_*}-\dfrac{U_m}{u_*} - E(-\dfrac{z}{L})^{-5/3} - G(-\dfrac{z}{L})^{-3}$,
    is plotted on the vertical axis. The slopes of the planar fit yield the coefficients $A$ and $D$
    associated with the $X_1 = (-\dfrac{z}{L})^{-1/3}$ and $X_2 = \epsilon'_3(-\dfrac{z}{L})^{1/3}$ terms, respectively. Darker and lighter symbols are above and below the plane respectively.}
  \label{fig:planar_lfc}
\end{figure}

\subsection{Local-free-convection layer}\label{sec:lfc}

As discussed in §\ref{subsec:lfco}, the expansion coefficients that result from matching between
the outer and inner–outer layers are determined using an iterative regression procedure.
To stabilise the estimation and mitigate the effects of multicollinearity, ridge regression (\citealt{hoerl1970ridge})
is employed, with the regularisation parameter $\lambda$ varied over the range $10^{-5} \le \lambda \le 1$.
The optimal value of $\lambda$ is selected using the corner criterion, yielding $\lambda = 0.0196$.
The resulting expansion coefficients are
\[
A = -4.37, \quad E = -1.58, \quad D = 0.57, \quad G = -0.23.
\]
Figure~\ref{fig:planar_lfc} shows 
 the velocity defect minus the (higher-order) contributions from the terms containing $E$ and $G$, denoted by the symbol $Y$, plotted as a function of
the leading-order variable $X_1=(-z/L)^{-1/3}$ and the second-order variable $X_2=\epsilon_3'(-z/L)^{1/3}$.
The figure also illustrates the planar fitting procedure used in the regression.
The remaining defect $Y$ depends linearly on the leading-order
contribution proportional to $A$ and on the higher-order term proportional to $D$. Therefore the dependence can be represented as
a plane. The plane fits the data points well with little scatter,
 demonstrating the validity of the asymptotic expansions and the regression procedure.

\begin{figure}
  \centering
  \includegraphics[width=1\linewidth]{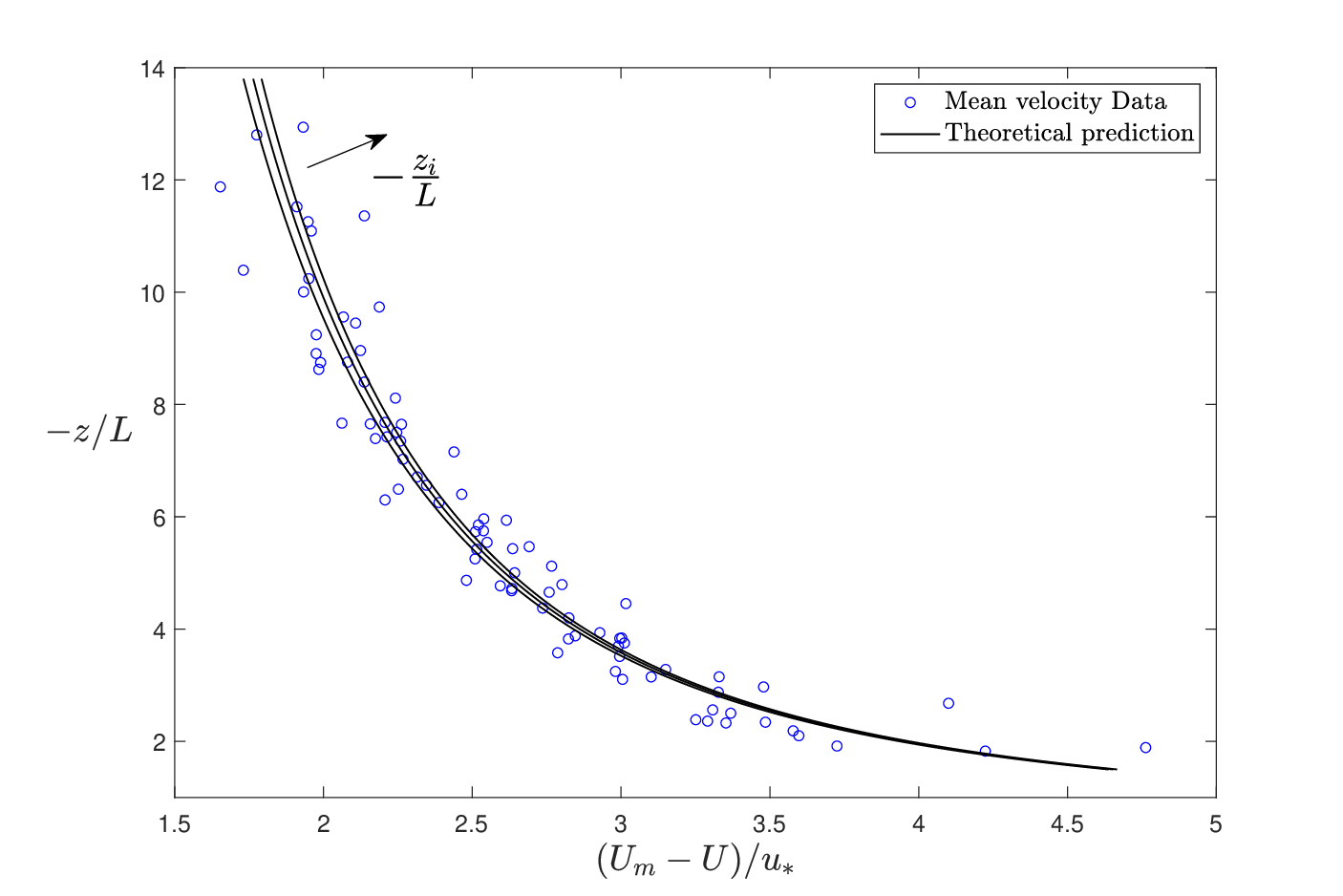}
  \caption{Mean velocity profiles (velocity defect) in the 
    local-free-convection layer. The measured data are shown as circles, and the black curves denote
    the predicted mean velocity profiles for a range of stability conditions characterized by different values of \(-z_i/L\): 43.6, 74.3 and 146.2.}
  \label{fig:lfc}
\end{figure}

The predicted velocity profiles in the local free-convection layer are presented for several
stationary periods in figure~\ref{fig:lfc}. The measured mean velocity is
shown as circles in terms of the non-dimensional velocity defect $(U_m - U)/u_*$, plotted
against the dimensionless height $-z/L$. The predicted profiles, shown as solid black lines, are obtained
from the asymptotic representation using the expansion coefficients determined from the matching procedure.
\begin{equation}
\frac{U - U_m}{u_*} = A(-z/L)^{-1/3} + E(-z/L)^{-5/3} + \epsilon'_3\,D(-z/L)^{1/3} + G(-z/L)^{-3},
\end{equation}
The leading-order term dominates the velocity defect over most of the local-free-convection layer and determines
the overall profiles. The higher-order term containing $E$ (asymptotically small for $-z/L\gg 1$), together with that containing $G$ (the next higher-order term),
acts predominantly at lower heights in the
inner–outer matching region, where departures from pure leading-order behaviour become significant.

\begin{figure}
  \centering
  \includegraphics[width=1\linewidth]{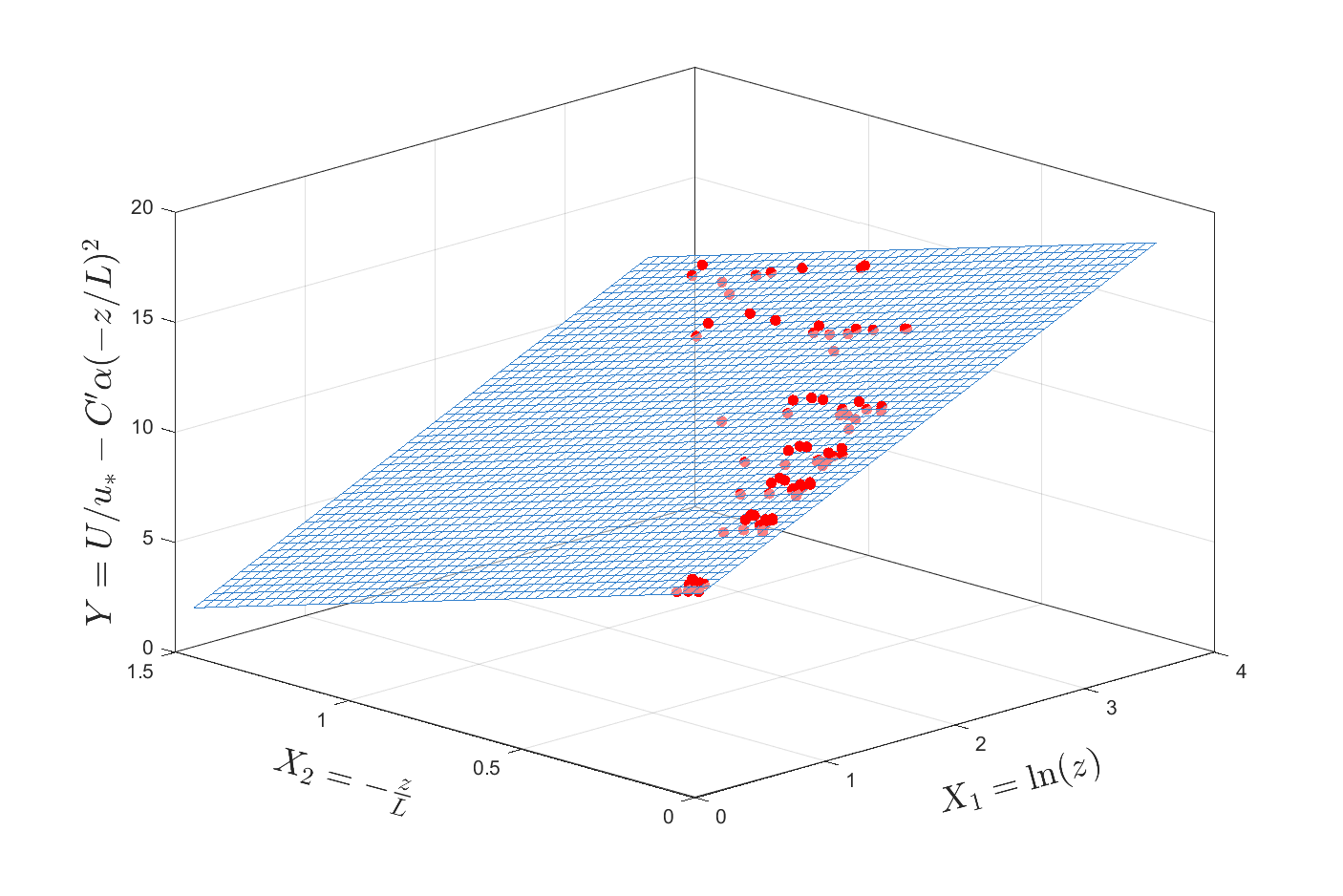}
  \caption{Planar fit of the velocity-profile in the log-law layer.
    The fit yields the inverse of the von K\'arm\'an constant and the coefficient of the first higher-order
     term, which accounts for a part of the deviations from the logarithmic velocity profile. Symbols same as in figure \ref{fig:planar_lfc}.}
  \label{fig:planar_log}
\end{figure}

The term with the coefficient $D$ contains the small perturbation parameter $\epsilon'_3$; therefore
its contribution is of higher-order for $-z/L \sim 1$, but increases with height. It is responsible
for the separation at large $-z/L$ of the predicted profiles for different $z_i/L$ values, which are
valid in the overlapping layer between the outer and inner–outer layers.
The close overall agreement between the asymptotic expansions and the observations for all the stationary
periods considered validates the prediction of the asymptotic analysis and confirms the robustness of the regression procedure.

Given the relatively large statistical uncertainties associated with the mean velocity profiles, it would not
be possible to accurately obtain the expansion coefficients if the expansion is used to fit individual profiles.
However, when the set of all the suitable profiles are used, the data size increases significantly, thereby resulting in
a more accurate regression. Similarly, comparisons of the asymptotic expansions to the measured mean velocity
should also be made by comparing the expansions to a set of profiles with different parameters, not to the individual profiles.

\subsection{Log law layer}

\begin{figure}
  \centering
  \includegraphics[width=1\linewidth]{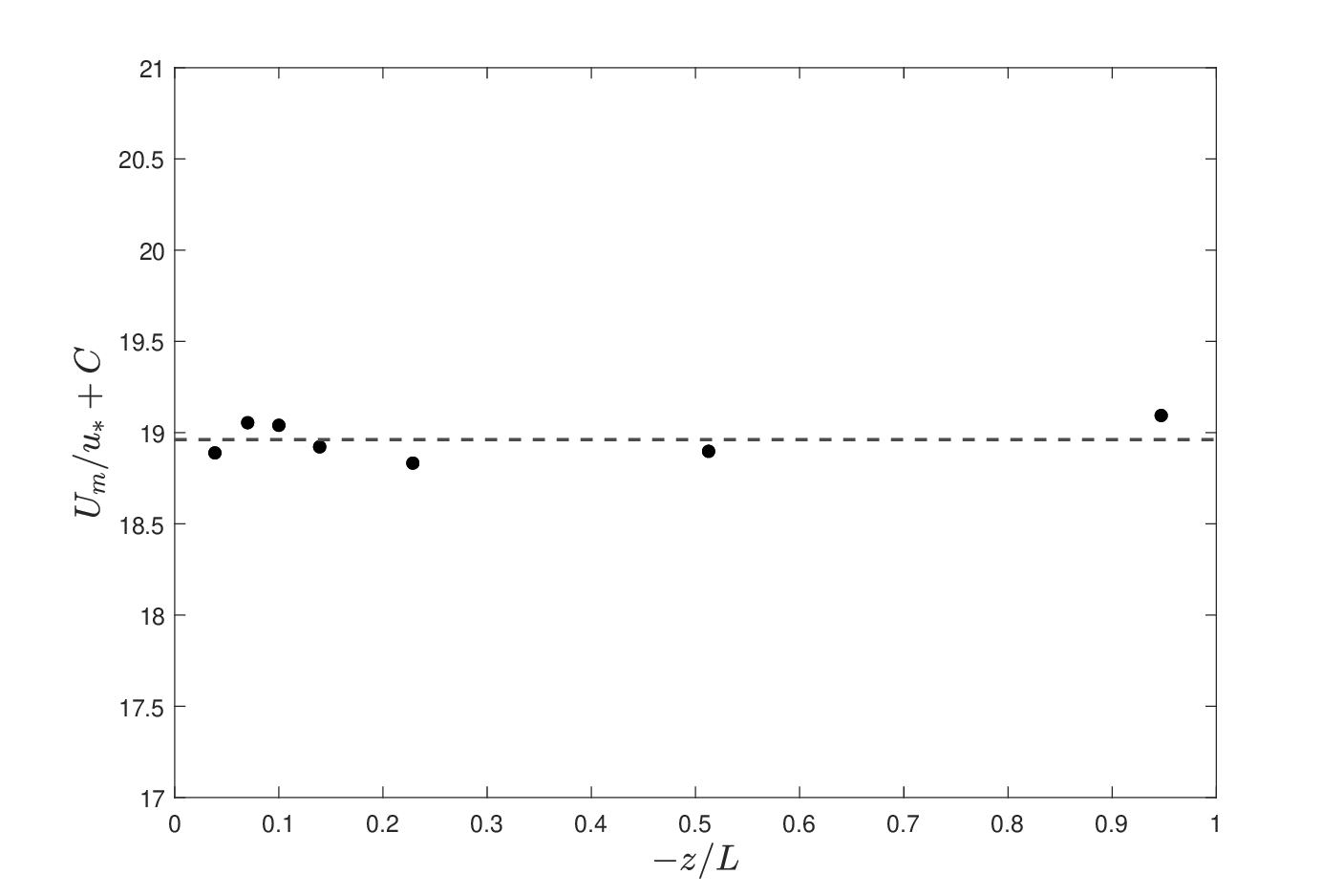}
  \caption{Non-dimensional mixed-layer mean velocity scale $U_m$ evaluated using data at different heights
    during the period 11:20–15:16 PDT on 26 August 2023.}
  \label{fig:Um_13}
\end{figure}

To determine the coefficients in the matching layer between the inner-outer and inner-inner layers, flux tower measurements of the mean velocity are
used. The regression is also carried out by representing the equation in the
form of a plane. 
Data points are considered for heights ranging from $z = 1\,\mathrm{m}$ up to $z = 1.3|L|$.
We find that within this range we are able to identify the leading-order (the log law) and the higher-order terms.
Figure~\ref{fig:planar_log} illustrates
the fitted plane obtained from the regression analysis.


The regression procedure yields the following coefficient values for the
surface-layer formulation. The value of the von Kármán constant, the stability correction parameters, and the
roughness height are obtained 
\begin{equation}\label{eq:kappa}
\kappa = 0.344, \qquad
C' = -4.841, \qquad
C'\alpha = 1.861, \qquad
h_0 = 0.045 \ \mathrm{m}.
\end{equation}
Again, given the uncertainties in the individual profiles, determination of the expansion coefficients with good accuracy is only possible using the set of
all the suitable profiles. 
The  roughness length obtained falls within
the range commonly reported for unstable atmospheric surface layer with similar terrain. The von K\'{a}rm\'{a}n constant is close that obtained by \cite{BWIB71}, but is
smaller than the values obtained in many field measurements (\citealt{Hogstrom96}). However, the previous values were invariably  obtained by fitting the log law, the leading-order profile,
to the measured profile, which includes contributions from the higher-order terms. 
 Since the coefficient for the second-order term, $C'$, is negative (equation (\ref{eq:kappa})),  these values tend to underestimate the leading-order term, and therefore overestimate
 $\kappa$. It would be useful to apply our procedure to a wider range of data to further investigate the issue of the value of the von K\'{a}rm\'{a}n constant.
\\



\begin{figure}
  \centering
  \includegraphics[width=1\linewidth]{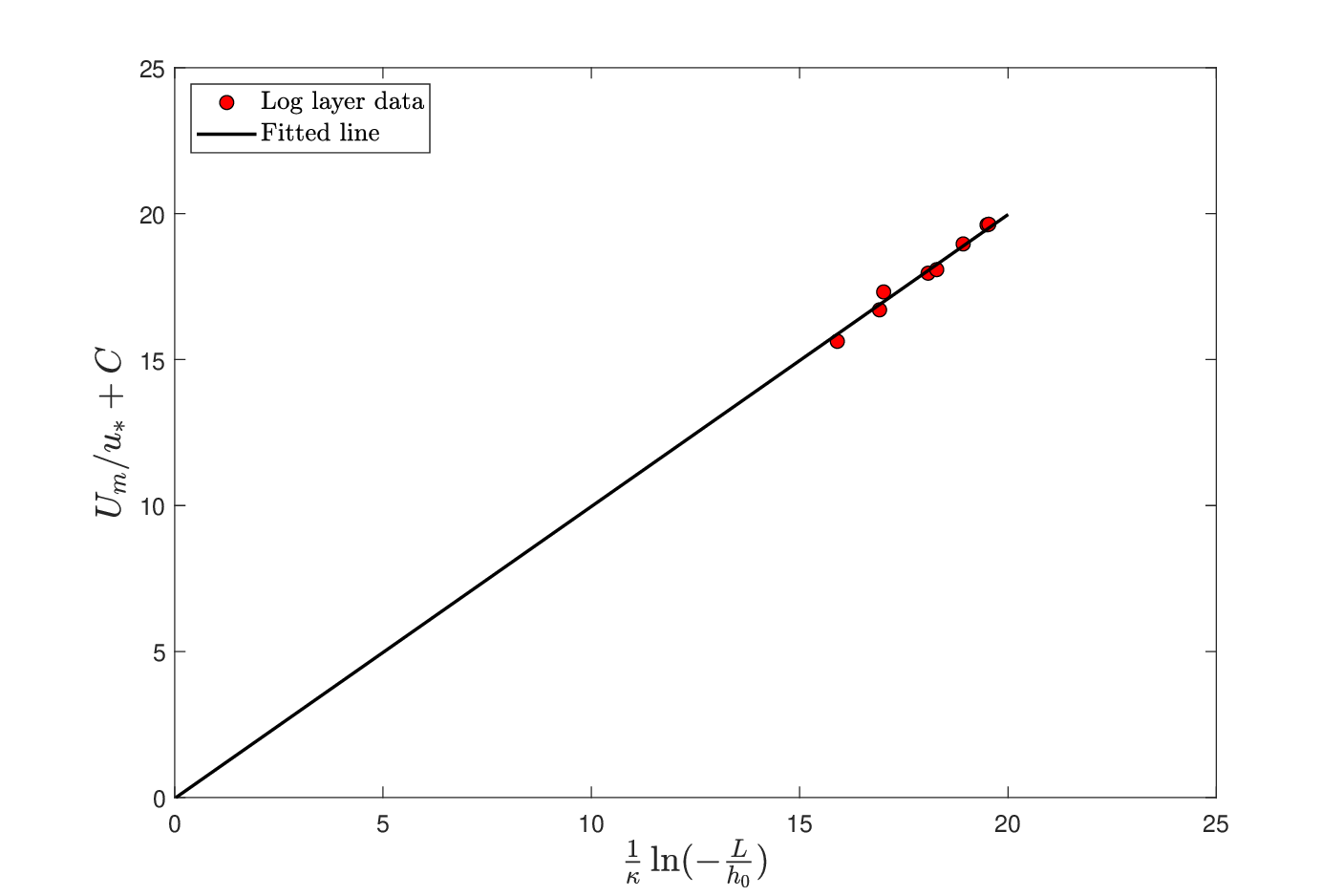}
  \caption{Comparison between the convective logarithmic friction law and data obtained in the log-law layer. The constant $C$ is combined with
  $U_m/u_*$. The line represents the friction law.}
  \label{fig:friction_law}
\end{figure}

\begin{figure}
  \centering
  \includegraphics[width=1\linewidth]{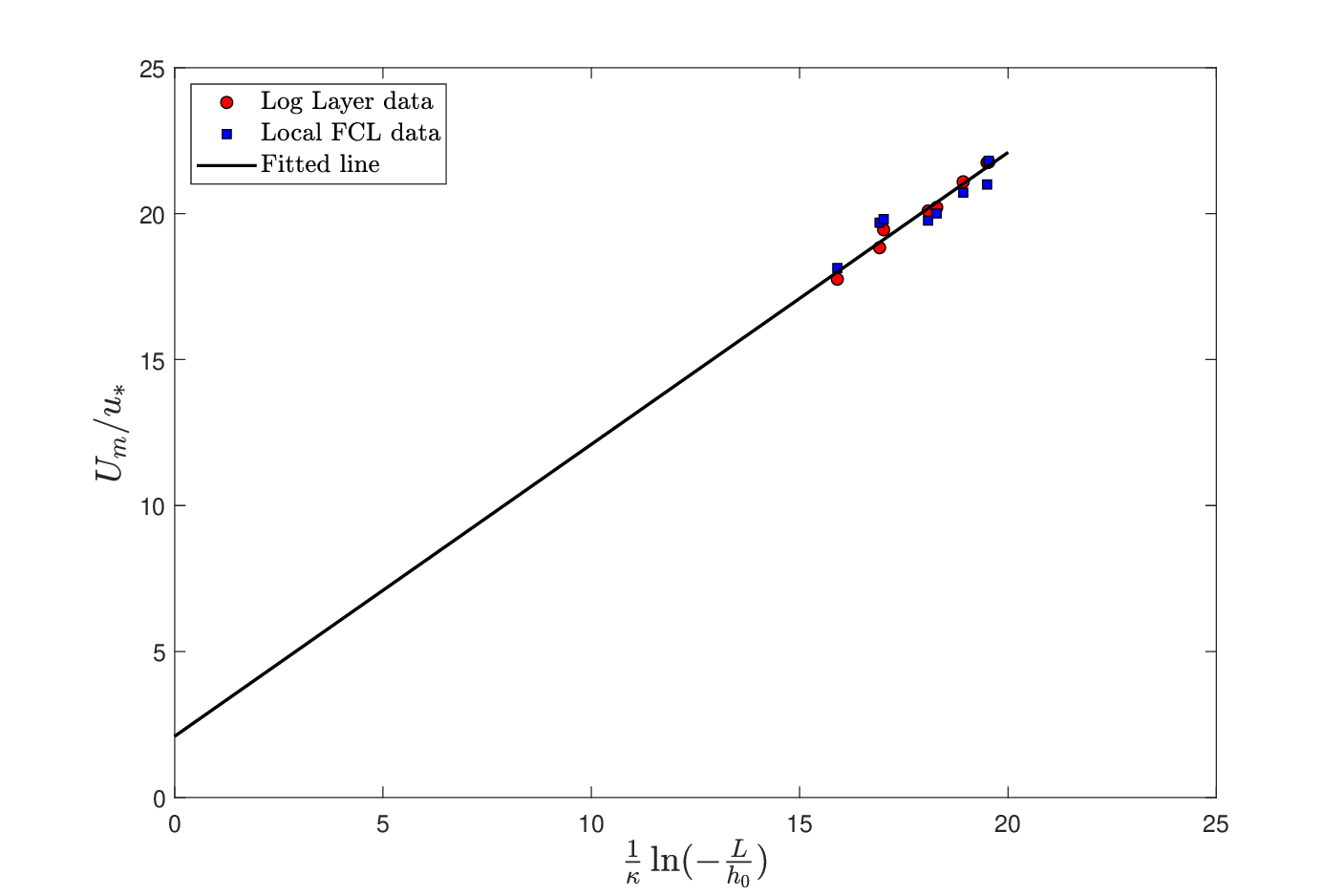}
  \caption{Comparison between the convective friction law and $U_m/u_*$  obtained using data in both matching layers. Circles and squares denote results from the log-law layer and
    the local-free-convection layer respectively.  The line represents the friction law.}
  \label{fig:friction_law_both}
\end{figure}

The matching results between the inner–inner and inner–outer layers can also be represented as
the surface-layer velocity defect
\begin{equation}
\frac{U - U_m}{u_*}
= \frac{1}{\kappa}\ln\!\left(-\frac{z}{L}\right)
+ C
+ C'\left[ \left(-\frac{z}{L}\right)
+ \alpha \left(-\frac{z}{L}\right)^2 \right],
\end{equation}
The expression can be rearranged as
\begin{equation}
\frac{U_m}{u_*} + C
= \frac{U}{u_*}
- \left[ \frac{1}{\kappa}\ln\!\left(-\frac{z}{L}\right)
+ C'\left(( -\frac{z}{L}) + \alpha \left(-\frac{z}{L}\right)^2 \right)\right].
\end{equation}
Here, the quantity $\dfrac{U_m}{u_*} + C$ represents an offset arising from
the matching condition between the inner-outer and inner-inner layers. For each velocity profile,
the right-hand side of the equation can be evaluated at all available heights,
yielding multiple estimates of this offset. An example for a single stationary period
is shown in figure~\ref{fig:Um_13}, where the evaluated values of $\dfrac{U_m}{u_*} + C$ are
plotted as a function of height. The values obtained at different heights show minimal variations,
indicating that the procedure provides an accurate estimate of the offset.
 The average of these values is therefore taken as the estimate of $\dfrac{U_m}{u_*} + C$ for
 that profile. This procedure provides an effective way to evaluate the constant $C$.


The offset, which contains the non-dimensional mixed-layer mean velocity scale, $U_m/u_\ast$, can be used
to validate the convective logarithmic friction law. In figure~\ref{fig:friction_law}, the values of
$U_m/u_\ast+C$ are plotted against $\tfrac{1}{\kappa}\ln\!\big(-L/h_0\big)$. The unity slope of the linear fit
indicates the validity of the friction law. The fact that different data points have different $z_i/L$ values
confirms that the validity of the friction law extends beyond the leading order.



Using the values of $U_m/u_\ast+C$ obtained from  matching  the inner-outer and inner-inner layers
for each profile, and the  values of $U_m/u_\ast$ (section \ref{sec:lfc}), the constant is determined to be $C=-2.13$.
The $U_m/u_\ast$ values obtained from both the local free-convection
layer and the log-law layer can also be used to validate the friction law (figure~\ref{fig:friction_law_both}).
The results  from both matching layers are nearly identical, demonstrating that the efficacy of the data analysis procedure. 



\begin{figure}
  \centering
  \includegraphics[width=1\linewidth]{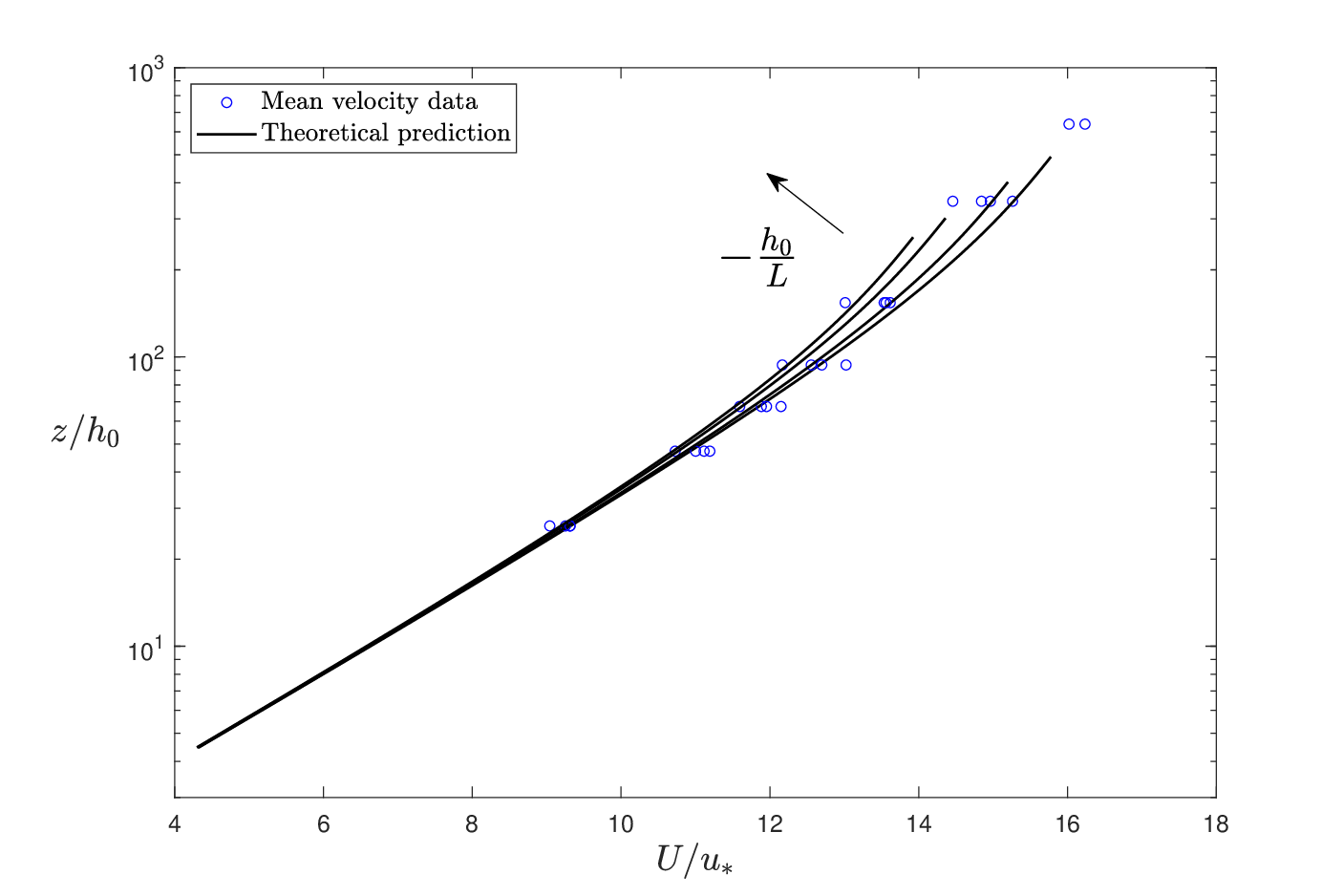}
  \caption{Mean velocity profiles in the matching layer between the Inner--outer
    and Inner--inner layers (Log-law layer). The measured data are plotted as blue circles,
    and the black curves denote the predicted mean velocity profiles for a range of
    stability conditions characterized by different values of $-\frac{h_0}{L}$: $(1.23,~ 1.49,~ 1.99,~ 2.31) \times 10^{-3}$}
  \label{fig:Velocity_Log}
 \end{figure}

The mean velocity profile near the surface, $U/u_\ast$, is plotted against
the non-dimensional height $z/h_0$  in figure~\ref{fig:Velocity_Log}. Very close to the surface, all curves nearly collapse,
indicating the dominance of the logarithmic term. As the height increases,
the curves begin to separate due to the influence of higher-order corrections associated with
the small non-dimensional parameter $\epsilon_3''=-\frac{h_0}{L}$, which accounts for the effects of the
buoyancy production. Although $\epsilon_3''$ is small, its contribution becomes increasingly
noticeable at larger $z/h_0$, allowing the expansions to better approximate the measured
velocity profile up to $-z/L\sim 1$. We note that for the data set used, $-z/L$ is generally larger than 0.1. As a result, one can hardly identify a logarithmic
profile, let alone determining the von K\'{a}rm\'{a}n constant. However, fitting the higher-order asymptotic expansion to the set of data 
as a whole allows us to accurately identify the log term, the von K\'{a}rm\'{a}n constant and the higher-order expansion coefficients.

\section{Conclusions}

We derived the higher-order mean velocity profile in the convective atmospheric boundary layer (CBL) using the method of matched asymptotic expansions
and obtained the expansion coefficients using the measurement data from the recent M$^2$HATS field campaign.

\cite{TD19b} have identified the three scaling layers, the mixed or outer layer, the Monin-Obukhov or inner-outer layer and the roughness or inner-inner layer.
In this work, for each layer a set of non-dimensional perturbation equations containing small (non-MOST) parameters was obtained  using the Reynolds-stress,
potential-temperature flux and
 potential temperature-variance budget equations and the mean momentum and  mean potential temperature equations.
The scaling of the terms in the budget equations were obtained using both the Monin-Obukhov similarity theory and the Multipoint
  Monin-Obukhov similarity theory. 
  The  small parameters with the most impact in the perturbation equations are $(-z_i/L)^{-4/3}$, $(-z_i/L)^{-2/3}$ and $-h_0/L$, where $z_i$, $L$ and $h_0$ are
  the inversion height, the Obukhov length and the roughness height respectively. These parameters represent the effects of the
  mean shear production of the $u$-variance, the unsteadiness of the inner-outer layer and the buoyancy production of the $w$-variance on  the mean profile.
The portion of the mean profile in each scaling layer is expanded asymptotically in terms of these small parameters. 

 \cite{TD19b}  have  performed asymptotic matching in the two overlapping layers between the three scaling layers, and obtained the leading-order
  expansions (the local-free-convection scaling and the log law).
  In the present work, the asymptotic matching between the outer and inner-outer layers to the second order led to the higher-order expansion terms,
  which account for the deviations from the local-free-convection scaling due to mean shear production of the $u$-variance and the unsteadiness,
  a result of finite values of $z_i/L$.
  The asymptotic matching between the inner-outer and inner-inner layers  results in the higher-order
  expansions that account for the deviations  from the log law due to buoyancy production, a result of finite values of $h_0/L$.

  The asymptotic expansions were used to fit (through linear regression) the set of all the suitable field data, which includes a range of $z_i/L$ and $h_0/L$ values,
  instead of individual profiles, to obtain the value of $U_m/u_*$ for each profile and the universal expansion coefficients. Using the data set increases the amount of
  regression data  compared with individual profiles, thereby reducing the  uncertainties in the regression results. 
  The theoretical  prediction shows an excellent agreement with the data set, best seen in figures \ref{fig:planar_lfc} and \ref{fig:planar_log}, indicating that the higher-order effects, i.e. the
  departures from the leading-order profile are well captured by the higher-order expansions.

  We note that because the $-z/L$ values for the profiles are generally larger than 0.1,  the leading-order logarithmic profile is hardly directly observable, making
  a determination of the  von K\'{a}rm\'{a}n constant very difficult. By fitting the higher-order expansions to
  the data set, the leading-order profile and the von K\'{a}rm\'{a}n constant can be indirectly obtained. 
The von K\'{a}rm\'{a}n constant obtained in the present work (0.344) is close that obtained by \cite{BWIB71} (0.35), but is
smaller than the values obtained in many field measurements (\citealt{Hogstrom96}). However, the previous values were invariably  obtained by fitting  the leading-order profile (the log law)
to the measured profile, the latter including contributions from the higher-order terms. 
 Since the second-order term is negative (equation (\ref{eq:kappa})),  such a fit tends to underestimate the leading-order term, and therefore overestimate
 $\kappa$. It would be useful to apply our procedure to a wider range of data to further investigate the issue of the value of the von K\'{a}rm\'{a}n constant.

 The asymptotic expansions also show that the convective logarithmic friction law derived by \cite{TD19b} is valid to at least second order.
 Therefore, $U_m/u_*$ only depends on $L/h_0$ and there are no other parameters involved.
The friction law shows very good agreement with the data. In contrast, in channel and pipe flows with smooth walls, the logarithmic frictional law is
a leading-order approximation (\citealt{Tennekes68,Afzal1976}).

The method of asymptotic expansion quantifies the dependencies of the mean velocity profile on the perturbation parameters and identifies the physical origin of the dependencies,
demonstrating its effectiveness as an approach to investigating the ABL. The higher-order mean velocity profile obtained in the present work can achieve higher accuracy
over pure empirical profiles obtained through data fitting alone, and could potentially benefit a wide range of applications.
  
\appendix

\section{Cross-order matching between the adjacent layers}\label{app:cross}

Here we provide the details of the cross-order matching between the outer and inner-outer layers.
Matching the second-order outer expansion with the third-order inner-outer expansion, we have
\begin{equation}
\epsilon_1 U_{o,2} = \epsilon_1' \epsilon_3' U_{io,31}. \text{ Thus } \epsilon_1 U_{o,2} = \epsilon_1 z_{o}^\alpha = \frac{\epsilon_1}{\epsilon_1'\epsilon_3'} \epsilon_1' \epsilon_3'z_{io}^\alpha(\frac{L}{z_i})^\alpha, \text{ requiring }
\end{equation}
\begin{equation}
 \frac{\epsilon_1}{\epsilon_1'\epsilon_3'}(\frac{L}{z_i})^\alpha = 1, \ \frac{u_*}{w_*} (\frac{z_i}{L})^{2/3-\alpha} = 1, \ \alpha = \frac{1}{3}.
\end{equation}
\begin{equation}
U_{o,22} = D(\frac{z}{z_i})^{1/3},\ 
U_{io,312} = D(-\frac{z}{L})^{1/3},\
\end{equation}
Similarly, we match other terms of different orders,
\begin{equation}
\epsilon_1 \epsilon_3 U_{o,31} = \epsilon_1' U_{io,2}
\end{equation}
\begin{equation}
\epsilon_1 \epsilon_3 U_{o,31} = \epsilon_1 \epsilon_3 z_o^\beta = \frac{\epsilon_1 \epsilon_3}{\epsilon_1'} \epsilon_1' z_{io}^\beta(\frac{L}{z_i})^\beta
\Rightarrow \frac{\epsilon_1 \epsilon_3}{\epsilon_1'}(\frac{L}{z_i})^\beta = 1, \frac{u_*}{w_*} (\frac{z_i}{L})^{-4/3-\beta} = 1 \Rightarrow \beta = -5/3.
\end{equation}
\begin{equation}
U_{o,313} = E(\frac{z}{z_i})^{-5/3},\ 
U_{io,22} = E(-\frac{z}{L})^{-5/3};\ 
\end{equation}

\begin{equation}
\epsilon_1 U_{o,2} = \epsilon_5'\epsilon_1' U_{io,31}
\end{equation}
\begin{equation}
\epsilon_5'\epsilon_1' U_{io,31} = \epsilon_5'\epsilon_1' z_{io}^\gamma = \frac{\epsilon_5'\epsilon_1'}{\epsilon_1}\epsilon_1 z_o^\gamma (\frac{z_i}{L})^\gamma \Rightarrow \frac{\epsilon_5'\epsilon_1'}{\epsilon_1}(\frac{z_i}{L})^\gamma = 1, \frac{w_*}{u_*}(\frac{z_i}{L})^{-1+\gamma} = 1  \Rightarrow \gamma = 2/3.
\end{equation}
\begin{equation}
U_{o,23} = F(\frac{z}{z_i})^{2/3},\
U_{io,322} = F(-\frac{z}{L})^{2/3},\
\end{equation}

\begin{equation}
\epsilon_1 \epsilon_3^2 U_{o,41} = \epsilon_1' U_{io,23}
\end{equation}
\begin{equation}
\begin{split}
  \epsilon_1 \epsilon_3^2 U_{o,41} &= \epsilon_1 \epsilon_3^2 z_o^\lambda = \frac{\epsilon_1 \epsilon_3^2}{\epsilon_1'}\epsilon_1' z_{io}^\lambda (-\frac{L}{z_i})^\lambda \Rightarrow \frac{\epsilon_1 \epsilon_3^2}{\epsilon_1'}(-\frac{L}{z_i})^\lambda = 1, \frac{u_*}{w_*}(-\frac{z_i}{L})^{-8/3-\lambda} = 1 \\
  &\quad \Rightarrow \lambda = -3.
 \end{split}
\end{equation}
\begin{equation}
U_{o,41} = G(\frac{z}{z_i})^{-3},\
U_{io,23} = G(-\frac{z}{L})^{-3};\
\end{equation}

\begin{equation}
\epsilon_1 \epsilon_3^2 U_{o,42} = \epsilon_5'\epsilon_1' U_{io,313}
\end{equation}
\begin{equation}
\begin{split}
  \epsilon_1 \epsilon_3^2 U_{o,42} &= \epsilon_1 \epsilon_3^2 z_o^\eta = \frac{\epsilon_1 \epsilon_3^2}{\epsilon_5'\epsilon_1'}\epsilon_5'\epsilon_1' z_{io}^\eta (-\frac{L}{z_i})^\eta \Rightarrow \frac{\epsilon_1 \epsilon_3^2}{\epsilon_5'\epsilon_1'}(-\frac{L}{z_i})^\eta = 1, \frac{u_*}{w_*}(-\frac{z_i}{L})^{-5/3-\eta} = 1 \\
 &\quad \Rightarrow \eta = -2.
\end{split}
\end{equation}
\begin{equation}
U_{o,42} = H(\frac{z}{z_i})^{-2},\
U_{io,323} = H(-\frac{z}{L})^{-2};\
\end{equation}

\begin{equation}
\epsilon_1 \epsilon_3^2 U_{o,43} = \epsilon_1'\epsilon_3' U_{io,323}
\end{equation}
\begin{equation}
\begin{split}
  \epsilon_1 \epsilon_3^2 U_{o,43} &= \epsilon_1 \epsilon_3^2 z_o^\xi = \frac{\epsilon_1 \epsilon_3^2}{\epsilon_1'\epsilon_3'}\epsilon_1'\epsilon_3' z_{io}^\xi (-\frac{L}{z_i})^\xi \Rightarrow \frac{\epsilon_1 \epsilon_3^2}{\epsilon_1'\epsilon_3'}(-\frac{L}{z_i})^\xi = 1, \frac{u_*}{w_*}(-\frac{z_i}{L})^{-2-\xi} = 1 \\
  &\quad \Rightarrow \xi = -7/3.
\end{split}
\end{equation}
\begin{equation}
U_{o,43} = I(\frac{z}{z_i})^{-7/3},\
U_{io,313} = I(-\frac{z}{L})^{-7/3}.\
\end{equation}

 In the similar way, the cross-order matching between the inner--outer and inner--inner layers can be carried out,
\begin{equation}
  \epsilon_1'U_{io,2} = \epsilon_1' \epsilon_3''U_{ii,32},
\end{equation}
\begin{equation}
  \begin{split}
    \text{Thus} \ \  \epsilon_1' U_{io,2} &= \epsilon_1 z_{io}^\alpha = \frac{\epsilon_1'}{\epsilon_1'\epsilon_3''}\epsilon_1'\epsilon_3'' z_{ii}^\alpha (-\frac{h_0}{L})^\alpha \\
    &\frac{\epsilon_1'}{\epsilon_1'\epsilon_3''}(-\frac{h_0}{L})^\alpha = 1,\ \ (-\frac{L}{h_0})(-\frac{h_0}{L})^\alpha = 1,  \ \ \alpha = 1.
  \end{split}
\end{equation}
\begin{equation}
  U_{io,22} = C'(-\frac{z}{L}),\
  U_{ii,32} = C'(\frac{z}{h_0});\
\end{equation}

\begin{equation}
  \epsilon_1'U_{io,2} = \epsilon_1' \epsilon_3''^2 U_{ii,41}
\end{equation}
\begin{equation}
  \begin{split}
     \epsilon_1' U_{io,2} &= \epsilon_1 z_{io}^\beta = \frac{\epsilon_1'}{\epsilon_1'\epsilon_3''^2}\epsilon_1'\epsilon_3''^2 z_{ii}^\beta (-\frac{h_0}{L})^\beta \Rightarrow \frac{\epsilon_1'}{\epsilon_1'\epsilon_3''^2}(-\frac{h_0}{L})^\beta = 1, (-\frac{L}{h_0})^2(-\frac{h_0}{L})^\beta = 1 \\
       &\quad \Rightarrow \beta = 2.
  \end{split}
\end{equation}
\begin{equation}
  U_{io,23} = E'(-\frac{z}{L})^2,\
  U_{ii,41} = E'(\frac{z}{h_0})^2.\
\end{equation}

\section{Iterative procedure for determining the expansion coefficients}\label{app:lfco}

Here we describe the procedure for determining the expansion coefficients for the leading-order term and three higher-order terms for
local-free-convection layer,
\begin{align}
\begin{split}
  U_o &= 1 + \epsilon_1 A(\frac{z}{z_i})^{-1/3} + \epsilon_1 D(\frac{z}{z_i})^{1/3} + \epsilon_1 \epsilon_3 E(\frac{z}{z_i})^{-5/3} + \epsilon_1 \epsilon_3^2 G(\frac{z}{z_i})^{-3},\\
\end{split}\\
\begin{split}
  U_{io} &= 1 + \epsilon_1' A(-\frac{z}{L})^{-1/3} + \epsilon_1' E(-\frac{z}{L})^{-5/3} + \epsilon_1' G(-\frac{z}{L})^{-3} + \epsilon_1' \epsilon_3' D (-\frac{z}{L})^{1/3},\\
\end{split}
\end{align}
The second equation can be written as 
\begin{equation}
    \frac{U}{u_*} = \frac{U_m}{u_*} + A(-\frac{z}{L})^{-1/3} + E(-\frac{z}{L})^{-5/3} + G(-\frac{z}{L})^{-3} + \epsilon_3' D (-\frac{z}{L})^{1/3},
     \label{eq:fcl}
\end{equation}
The coefficients $A$, $E$, $D$, $G$, which are identical for all profiles, and the offset $U_m/u_\ast$ for each profile are determined through an iterative multi-step procedure. 

We first write the mixed-layer defect to include only one higher-order term
\begin{equation}
\frac{U}{u_\ast} = \frac{U_m}{u_\ast} + A(-z/L)^{-1/3} + E(-z/L)^{-5/3},
\end{equation}
For each profile, a planar fit is performed using the liner equation
\begin{equation}
Y = \frac{U_m}{u_\ast} + A X_1 + E X_2,
\end{equation}
where $Y = U/u_\ast$, $X_1 = (-z/L)^{-1/3}$, and $X_2 = (-z/L)^{-5/3}$. The intercept from this fit provides
an updated $U_m/u_\ast$ value for the profile.

In the second step, all the profiles are combined and a least-squares fit of the following equation is performed,
\begin{equation}
\frac{U}{u_\ast} - \frac{U_m}{u_\ast} = A(-z/L)^{-1/3} + E(-z/L)^{-5/3} + D \, \epsilon_3' (-z/L)^{1/3} + G(-z/L)^{-3},
\end{equation}
By setting $Y = U/u_\ast - U_m /u_\ast$, $X_1 = (-z/L)^{-1/3}$, $X_2 = (-z/L)^{-5/3}$, $X_3 = \epsilon_3' (-z/L)^{1/3}$, and $X_4 = (-z/L)^{-3}$, a
 linear regression can be performed,  providing estimates of the coefficients $A$, $E$, $D$, $G$.

 In the third step, the new values of the coefficients are used to update the $U_m/u_\ast$ value for each profile.
 For up to the 10th iteration, $U_m/u_\ast$ is updated as
\begin{equation}
\frac{U_m}{u_\ast} = \frac{U}{u_\ast} - A(-z/L)^{-1/3} - E(-z/L)^{-5/3}.
\end{equation}
For iterations beyond the 10th, the $D$ and $G$ terms are also included to improve the accuracy
\begin{equation}
\frac{U_m}{u_\ast} = \frac{U}{u_\ast} - A(-z/L)^{-1/3} - E(-z/L)^{-5/3} - D \, \epsilon_3' (-z/L)^{1/3} - G(-z/L)^{-3}.
\end{equation}
Steps two and three are repeated iteratively until the value of $U_m/u_\ast$ for each profile
converges with the difference between iterations less than $10^{-6}$.

\begin{figure}
  \centering
  \includegraphics[width=1\linewidth]{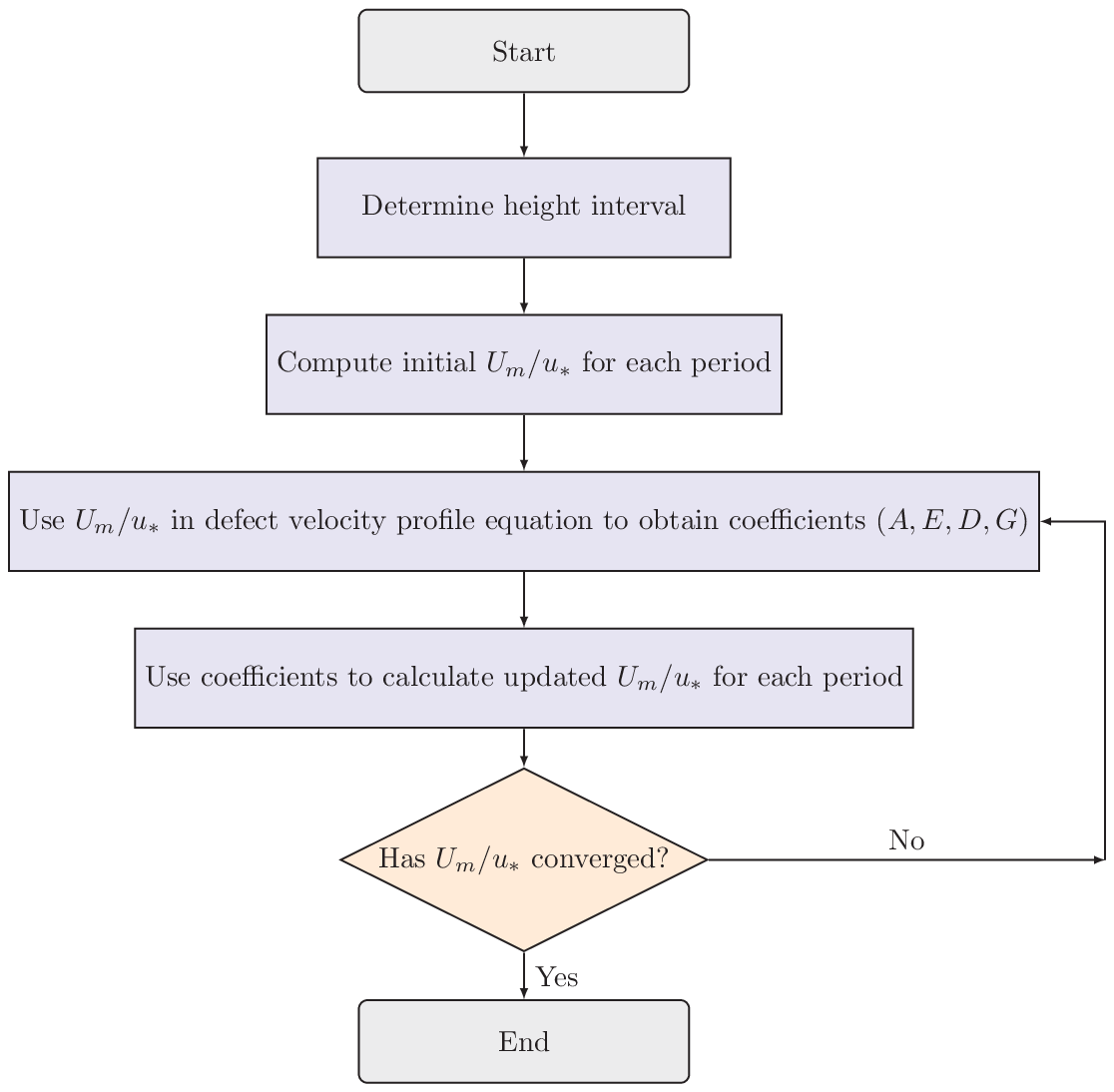}
  \caption{Flowchart of the iterative procedure for determining the coefficients in the Local Free Convection layer}
  \label{fig:flowchart}
\end{figure}

The flowchart in figure~\ref{fig:flowchart} depicts the iterative procedure used to determine the coefficients.

\section{Ridge Regression for Stabilizing Coefficient Estimation}\label{app:ridge}

During the above regression process, it is observed that the parameter $D$ tends to blow up  when the higher-order
term $(-z/L)^{1/3}$ is small, making the regression problem ill-conditioned. As a result, the least-squares
fit may overemphasize small fluctuations in the data by producing unrealistically large values of $D$. To mitigate
this instability, we employed the ridge regression method, which penalizes large coefficient magnitudes.
Specifically, the ridge regression method minimizes the following cost function
\begin{equation}
    J(\boldsymbol{\beta}) = \| X\boldsymbol{\beta} - Y \|_2^2 + \lambda \| \boldsymbol{\beta} \|_2^2
\end{equation}
where 
\[
X = \left[ (-z/L)^{-1/3}, \, (-z/L)^{-5/3}, \, (-z/L)^{1/3}, \, (-z/L)^{-3}, \, 1 \right]
\]
is the data matrix,
\[
\boldsymbol{\beta} = [A,\, E,\, D,\, G, \, \text{constant}]^T
\]
is the coefficient column matrix, 
\[
Y = \frac{U}{u_*} - \frac{U_m}{u_*}
\]
is the response column matrix and $\lambda$ is the regularization parameter controlling the trade-off between
the bias  and uncertainties of the estimated coefficients. The analytical solution is given by
\begin{equation}
    \boldsymbol{\beta}_{\text{ridge}} = (X^T X + \lambda I)^{-1} X^T Y
\end{equation}
where $I$ is the identity matrix.

\begin{figure}
  \centering
  \includegraphics[width=1\linewidth]{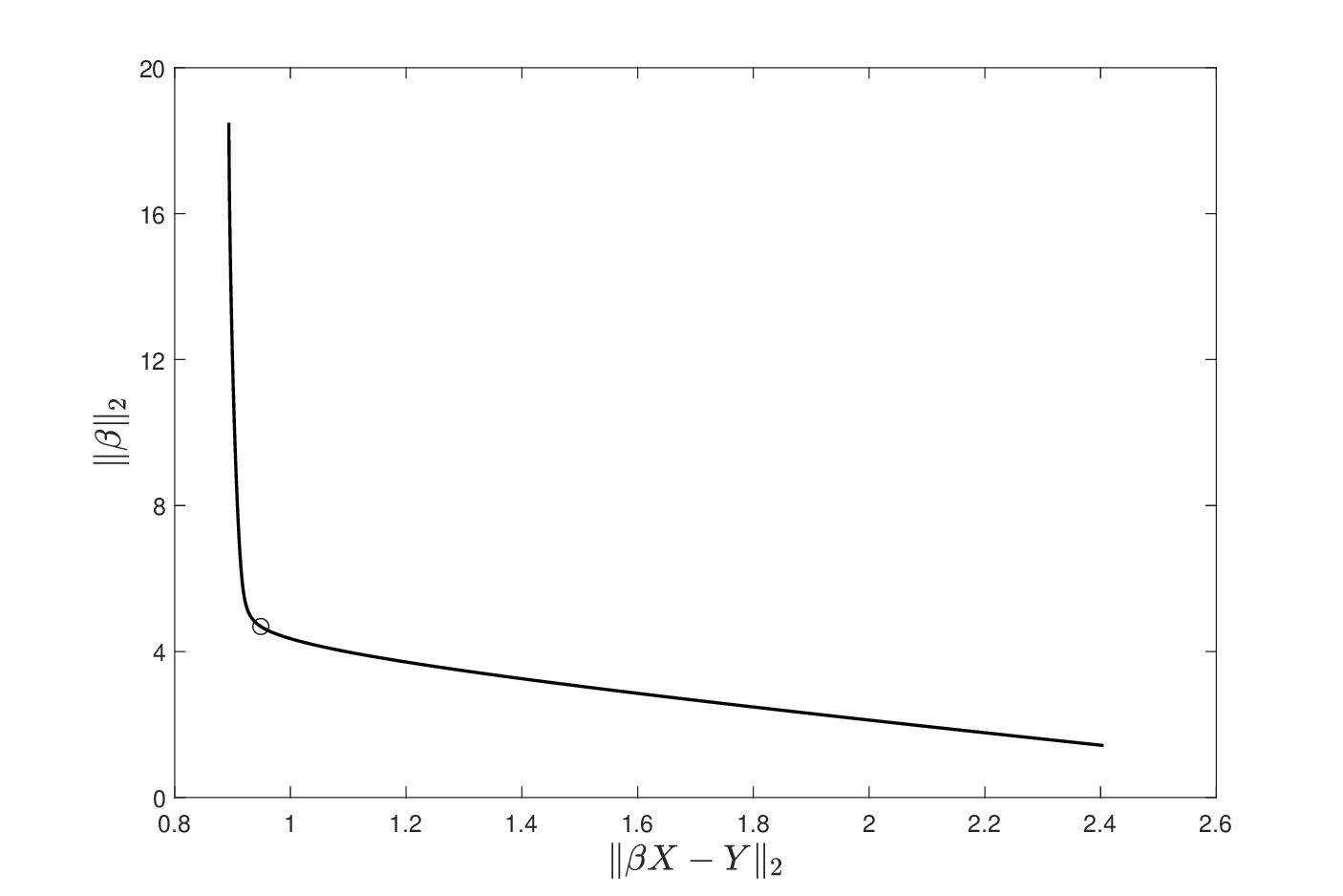}
  \caption{L--curve model to determine stabilized \(\lambda\)}
  \label{fig:Lcurve}
\end{figure}

To determine the optimal value of $\lambda$, we apply the L--curve method, which
provides a graphical criterion for optimization.
The L-curve is a plot of the solution norm $\|\boldsymbol{\beta}\|_2$ (vertical axis)
versus the residual norm $\|X\boldsymbol{\beta} - Y\|_2$ (horizontal axis) for a wide range of $\lambda$ values (figure~\ref{fig:Lcurve}).

For small $\lambda$, the regularization is weak, producing a small residual normal define, but a large
coefficient norm (due to over-fitting) and larger bias. For large $\lambda$, the regularization dominates,
yielding small a coefficient norm but large residual (due to over-smoothing) and larger uncertainties. The corner of the
L-curve, where the curvature is maximum, represents the point of optimal compromise between bias and uncertainty. 
(Increasing $\lambda$ further would not significantly reduce the bias but would
noticeably increase the uncertainty. On the other hand, reducing  $\lambda$ would not significantly reduce the uncertainty but would
significantly increase the bias). Therefore the corner 
value is selected as the optimal value of the regularization parameter.

\section{Bootstrap-Based Uncertainty Estimation}\label{app:Bootstrap}

To assess the statistical uncertainty of the expansion coefficients $A$, $E$, $D$, etc.~obtained
through regression, we
use a bootstrap resampling approach, which does not require an assumption on any particular uncertainty distribution. 
The approach  repeatedly draw samples with replacement
from the observed data and recomputes the coefficients for each resampled dataset. The population of the recomputed coefficients are then used
(as if they were truly computed from independent datasets) to 
approximate the sample distribution of the estimators (\citealt{Hall90}).
In the present work, a total of 2000 resampled datasets are generated. To ensure that each resampled dataset adequately
represented the full range of atmospheric stability conditions, the original dataset is first partitioned
into three subsets based on the $z_i/L$.
(i) low $z_i/L$ (weekly convective or near-neutral), (ii) mid $z_i/L$ (moderately convective), and (iii) high $z_i/L$ (strongly convective) regimes. Samples are drawn with replacement
from each subset and combined to form a dataset spanning all the stability ranges.
 This resulted in an ensemble of 2000 samples for each expansion coefficient.
 The associated uncertainty is quantified by the standard deviation.
The bootstrap approach ensures that the estimated uncertainties reflect the combined effects of measurement noise,
sensitivity of the fitting procedure, and variability across atmospheric stability ranges.

\begin{figure}
  \centering
  \includegraphics[width=6in]{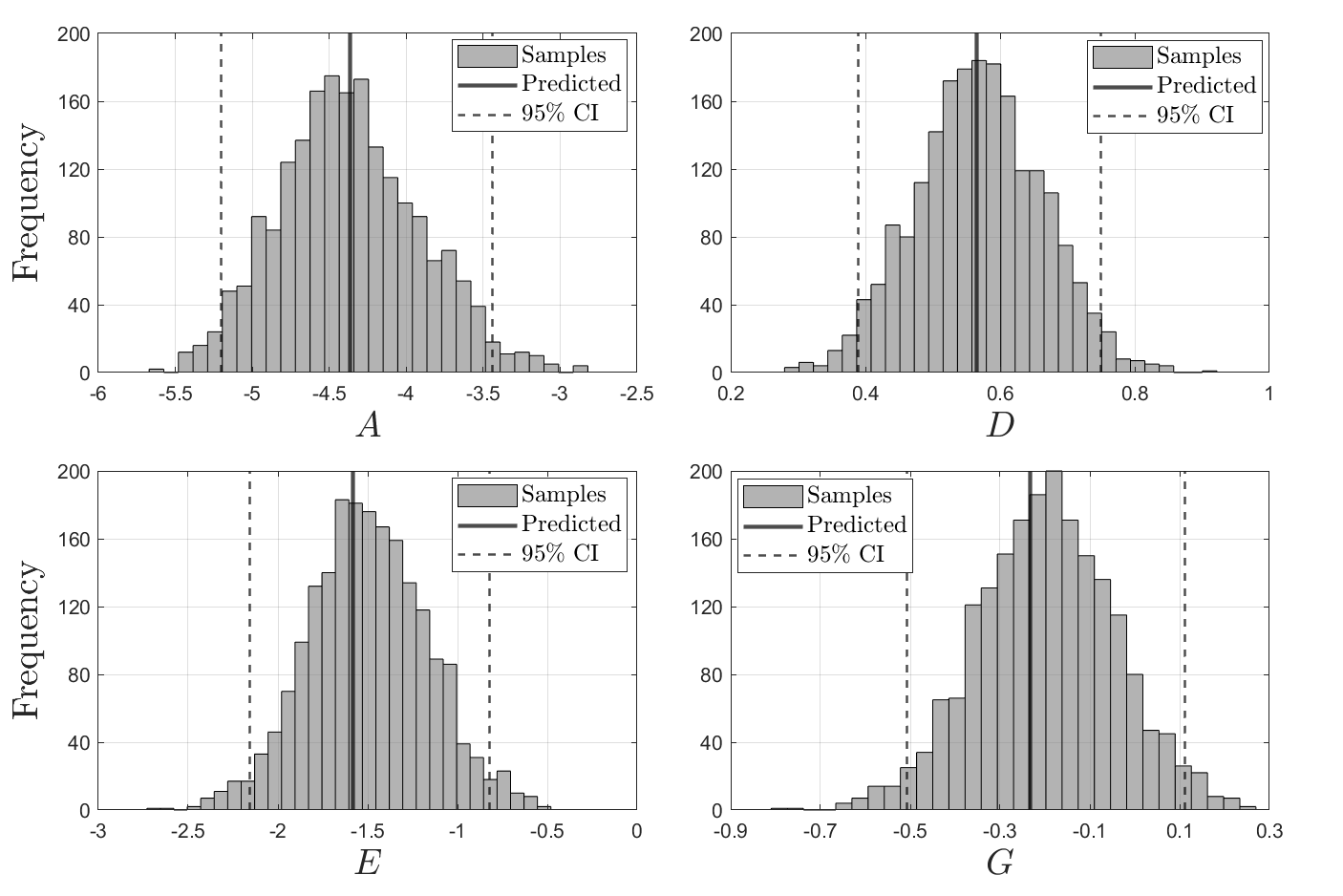}
  \caption{Uncertainty of the coefficients through bootstrap resampling for the local-free-convection layer}
  \label{fig:boot_FCL}
\end{figure}
\begin{figure}
  \centering
  \includegraphics[width=1\linewidth]{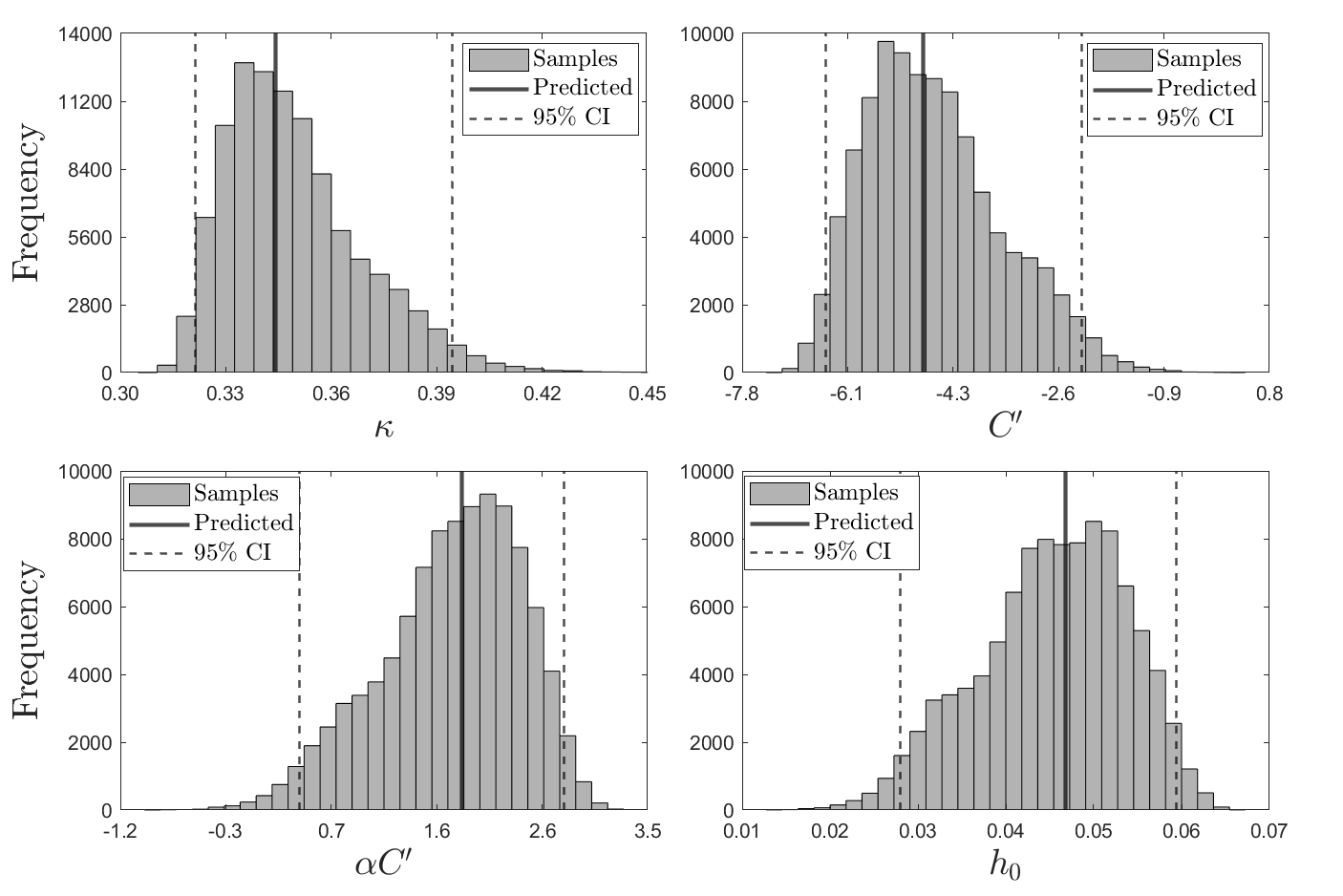}
  \caption{Uncertainty of the coefficients through bootstrap resampling for the Log law layer}
  \label{fig:boot_Log}
\end{figure}

The resulting bootstrap distributions for the expansion coefficients in the local-free-convection layer
and the logarithmic layer are shown in figures~\ref{fig:boot_FCL} and~\ref{fig:boot_Log}, respectively.
In the local-free-convection layer, the histograms indicate that the bootstrap realizations are approximately
Gaussian. 
In contrast, the coefficient distributions associated with the logarithmic layer exhibit noticeably
skewed and stretched shapes. The bootstrap approach can provide a better estimate of the confidence intervals for
these asymmetric distributions.

\begin{table} 
\begin{center} 
\begin{tabular}{c  c c c c c c} 
\hline 
  Expansion coefficients of      & $A$   & $E$ & $D$ & $G$ \\ 
  local free convection layer     &  & & & &  &  \\
                             &  & & & &  &  \\
 $\sigma$         &$0.46$  &$0.33$ &$0.09$ &$0.16$ \\
                              &  & & & &  &  \\
 $ 95\%~\mathrm{CI}$      &$[-5.20,~ -3.44]$  &$[-2.16,~ -0.82]$ &$[0.39,~ 0.75]$ &$[-0.51,~ 0.11]$ \\
        &  & &  &  \\ 
\hline
  Expansion coefficients of        & $\kappa$   & $C'$ & $\alpha C'$ & $h_0$ \\ 
  log law layer              &  & & & &  &  \\
                                      &  & & & &  &  \\
  $\sigma$          &$0.02$  &$1.12$ &$0.63$ &$0.008$ \\
                    &  & & & &  &  \\
  $95\%~\mathrm{CI}$       & $[0.32,~ 0.39]$ &$[-6.42,~ -2.24]$ &$[0.40,~ 2.78]$ &$[0.03,~ 0.06]$ \\
   &   &  &  &  \\ 
\hline 
  
\end{tabular} 
\caption{Standard deviation and confidence intervals of the expansion coefficients and roughness height} 
\label{tab:std} 
\end{center} 
\end{table}


\vspace{48pt}
\bibliographystyle{jfm}

\end{document}